\documentclass[twoside,twocolumn,9pt]{article}
\usepackage{extsizes}
\usepackage[super,sort&compress,comma]{natbib} 
\usepackage[version=3]{mhchem}
\usepackage[left=1.5cm, right=1.5cm, top=1.785cm, bottom=2.0cm]{geometry}
\usepackage{balance}
\usepackage{mathptmx}
\usepackage{sectsty}
\usepackage{graphicx} 
\usepackage{lastpage}
\usepackage[format=plain,justification=justified,singlelinecheck=false,font={stretch=1.125,small,sf},labelfont=bf,labelsep=space]{caption}
\usepackage{float}
\usepackage{fancyhdr}
\usepackage{fnpos}
\usepackage[english]{babel}
\addto{\captionsenglish}{%
  
}
\usepackage{array}
\usepackage{droidsans}
\usepackage{charter}
\usepackage[T1]{fontenc}
\usepackage[usenames,dvipsnames]{xcolor}
\usepackage{setspace}
\usepackage[compact]{titlesec}
\usepackage{hyperref}
\usepackage{chemformula} % Formula subscripts using \ch{}
\usepackage{epstopdf}%This line makes .eps figures into .pdf - please comment out if not required.

\definecolor{cream}{RGB}{222,217,201}

\begin{document}

\pagestyle{fancy}
\thispagestyle{plain}
\fancypagestyle{plain}{
%%%HEADER%%%
\renewcommand{\headrulewidth}{0pt}
}
%%%END OF HEADER%%%

%%%PAGE SETUP - Please do not change any commands within this section%%%
\makeFNbottom
\makeatletter
\renewcommand\LARGE{\@setfontsize\LARGE{15pt}{17}}
\renewcommand\Large{\@setfontsize\Large{12pt}{14}}
\renewcommand\large{\@setfontsize\large{10pt}{12}}
\renewcommand\footnotesize{\@setfontsize\footnotesize{7pt}{10}}
\makeatother

\renewcommand{\thefootnote}{\fnsymbol{footnote}}
\renewcommand\footnoterule{\vspace*{1pt}% 
\color{cream}\hrule width 3.5in height 0.4pt \color{black}\vspace*{5pt}} 
\setcounter{secnumdepth}{5}

\makeatletter 
\renewcommand\@biblabel[1]{#1}            
\renewcommand\@makefntext[1]% 
{\noindent\makebox[0pt][r]{\@thefnmark\,}#1}
\makeatother 
\renewcommand{\figurename}{\small{Fig.}~}
\sectionfont{\sffamily\Large}
\subsectionfont{\normalsize}
\subsubsectionfont{\bf}
\setstretch{1.125} %In particular, please do not alter this line.
\setlength{\skip\footins}{0.8cm}
\setlength{\footnotesep}{0.25cm}
\setlength{\jot}{10pt}
\titlespacing*{\section}{0pt}{4pt}{4pt}
\titlespacing*{\subsection}{0pt}{15pt}{1pt}
%%%END OF PAGE SETUP%%%

%%%FOOTER%%%
\fancyfoot{}
\fancyfoot[LO,RE]{\vspace{-7.1pt}\includegraphics[height=9pt]{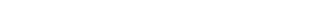}}
\fancyfoot[CO]{\vspace{-7.1pt}\hspace{11.9cm}\includegraphics{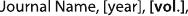}}
\fancyfoot[CE]{\vspace{-7.2pt}\hspace{-13.2cm}\includegraphics{head_foot/RF}}
\fancyfoot[RO]{\footnotesize{\sffamily{1--\pageref{LastPage} ~\textbar  \hspace{2pt}\thepage}}}
\fancyfoot[LE]{\footnotesize{\sffamily{\thepage~\textbar\hspace{4.65cm} 1--\pageref{LastPage}}}}
\fancyhead{}
\renewcommand{\headrulewidth}{0pt} 
\renewcommand{\footrulewidth}{0pt}
\setlength{\arrayrulewidth}{1pt}
\setlength{\columnsep}{6.5mm}
\setlength\bibsep{1pt}
%%%END OF FOOTER%%%

%%%FIGURE SETUP - please do not change any commands within this section%%%
\makeatletter 
\newlength{\figrulesep} 
\setlength{\figrulesep}{0.5\textfloatsep} 

\newcommand{\topfigrule}{\vspace*{-1pt}% 
\noindent{\color{cream}\rule[-\figrulesep]{\columnwidth}{1.5pt}} }

\newcommand{\botfigrule}{\vspace*{-2pt}% 
\noindent{\color{cream}\rule[\figrulesep]{\columnwidth}{1.5pt}} }

\newcommand{\dblfigrule}{\vspace*{-1pt}% 
\noindent{\color{cream}\rule[-\figrulesep]{\textwidth}{1.5pt}} }

\makeatother
%%%END OF FIGURE SETUP%%%

%%%TITLE, AUTHORS AND ABSTRACT%%%
\twocolumn[
  \begin{@twocolumnfalse}
{\includegraphics[height=30pt]{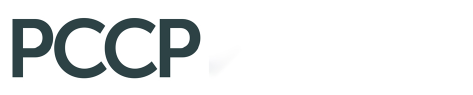}\hfill\raisebox{0pt}[0pt][0pt]{\includegraphics[height=55pt]{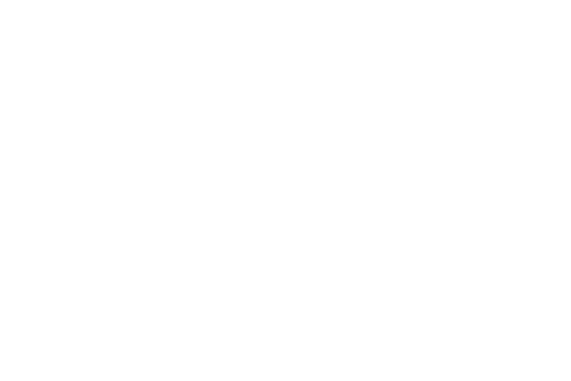}}\\[1ex]
\includegraphics[width=18.5cm]{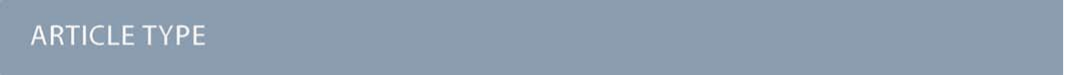}}\par
\vspace{1em}
\sffamily
\begin{tabular}{m{4.5cm} p{13.5cm} }

\includegraphics{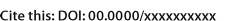} & \noindent\LARGE{\textbf{Formation of hydroxy, cyano and ethynyl derivatives of \ch{C4H4} isomers in the interstellar medium$^\dag$}} \\%Article title goes here instead of the text "This is the title"
\vspace{0.3cm} & \vspace{0.3cm} \\

 & \noindent\large{Mario Largo,\textit{$^{a}$} Miguel Sanz-Novo,\textit{$^{b}$} and Pilar Redondo$^{\ast}$\textit{$^{a}$}} \\%Author names go here instead of "Full name", etc.

\includegraphics{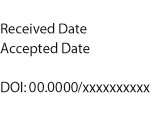} & \noindent\normalsize{The study of cyclic hydrocarbons is of utmost relevance in current astrochemical research, as they are considered to be among the most significant reservoirs of carbon in the interstellar medium. However, while unsaturated cyclic hydrocarbons with three, five, and six carbon atoms have been widely investigated, the highly strained antiaromatic cyclobutadiene (\ch{c-C4H4}) still remains uncharted. Here, we employed high-level CCSD(T)-F12/cc-pVTZ-F12//B2PLYPD3/aug-cc-pVTZ theoretical calculations to analyze whether the cyano (CN), ethynyl (CCH), and hydroxy (OH) derivatives of \ch{c-C4H4} and its structural isomers butatriene (\ch{H2CCCCH2}) and vinylacetylene (\ch{H2CCHCCH}) can readily form via the gas-phase reaction:  \ch{C4H4 + X → C4H3X + H} (where X = CN, CCH, and OH). For each system, we thoroughly explored the corresponding potential energy surfaces, identifying their critical points to enable a detailed analysis of the thermochemistry. Hence, we found various exothermic pathways for the formation of CN and CCH derivatives of butatriene and vinylacetylene, with no net activation barriers, while the formation of the OH derivatives is in general less favorable. Prior to the mechanistic study, we also analyzed the complete conformational panorama and stability of all the derivatives at the CCSD(T)-F12/cc-pVTZ-F12 level. Overall, \ch{c-C4H3CN} and \ch{c-C4H3CCH} emerge as particularly promising candidates for interstellar detection, provided that the parental \ch{c-C4H4} is present in the gas phase. These findings highlight the potential for detecting polar derivatives of \ch{c-C4H4} as indirect evidence of its presence in the ISM, as it appears to be "invisible" to radioastronomical observations. Also, this study underscores the need for future laboratory and theoretical efforts to characterize the spectroscopic properties of the proposed derivatives, paving the way for their eventual identification in space.} \\ 

\end{tabular}

 \end{@twocolumnfalse} \vspace{0.6cm}

  ]
%%%END OF TITLE, AUTHORS AND ABSTRACT%%%

%%%FONT SETUP - please do not change any commands within this section
\renewcommand*\rmdefault{bch}\normalfont\upshape
\rmfamily
\section*{}
\vspace{-1cm}

%%%FOOTNOTES%%%

\footnotetext{\textit{$^{a}$~Computational Chemistry Group, Departamento de Química Física y Química Inorgánica, Universidad de Valladolid, E-47011 Valladolid, Spain. E-mail: pilar.redondo@uva.es}}
\footnotetext{\textit{$^{b}$~Centro de Astrobiología (CAB), CSIC-INTA, Carretera de Ajalvir km 4, Torrejón de Ardoz, 28850 Madrid, Spain. }}

%Please use \dag to cite the ESI in the main text of the article.
%If you article does not have ESI please remove the the \dag symbol from the title and the footnotetext below.
\footnotetext{\dag~Supplementary Information available. See DOI: 10.1039/cXCP00000x/}
%: [details of any supplementary information available should be included here]
%additional addresses can be cited as above using the lower-case letters, c, d, e... If all authors are from the same address, no letter is required

%\footnotetext{\ddag~Additional footnotes to the title and authors can be included \textit{e.g.}\ `Present address:' or `These authors contributed equally to this work' as above using the symbols: \ddag, \textsection, and \P. Please place the appropriate symbol next to the author's name and include a \texttt{\textbackslash footnotetext} entry in the the correct place in the list.}

%%%END OF FOOTNOTES%%%

%%%MAIN TEXT%%%%
%The main text of the article\cite{Mena2000} should appear here.

%\subsection{This is the subsection heading style}
%Section headings can be typeset with and without numbers.\cite{Abernethy2003}

%\subsubsection{This is the subsubsection style.~~} These headings should end in a full point.  

%\paragraph{This is the next level heading.~~} For this level please use \texttt{\textbackslash paragraph}. These headings should also end in a full point.

\section{Introduction}
In recent times, among the wide variety of molecules identified to date in the interstellar medium (ISM; see \citealt{McGuire22census,Jimenez-Serra2025} for a recent census), cyclic hydrocarbons have drawn considerable attention from the astrophysical community. Particularly, the presence of benzene, the simplest aromatic unit, and diverse polycyclic aromatic hydrocarbons (PAHs) has been confirmed through the detection of infrared (IR) emission features in a variety of interstellar environments.\citep{tielens2008,garcia-bernete2022,chown2024_orion,cernicharo2001,tabone2023,arabhavi2024}\\
However, pure cyclic hydrocarbons often exhibit a very low or even zero permanent dipole moment, which makes them "invisible" to rotational spectroscopy, further hampering its radioastronomical identification. An exception to this trend is cyclopropenylidene ($c$-\ch{C3H2}; $\mu$$_b$ = 3.27 D), the first cyclic molecule identified in the ISM, \citep{Thaddeus:1985hw,Madden1989,Agundez:2013ga} as well as cyclopentadiene ($c$-\ch{C5H6}; $\mu$$_b$ = 0.416 D) and indene ($c$-\ch{C9H8}; $\mu$$_a$ = 0.50 D and $\mu$$_b$ = 0.37 D) \citep{cernicharo2021_indene}, which have been recently detected toward the dark cold pre-stellar core Taurus Molecular Cloud 1 (TMC-1). This source accounts for the majority of new interstellar detections on cylic systems, achieved during the course of two sensitive molecular line surveys: the Q-band Ultrasensitive Inspection Journey to the Obscure TMC-1 Environment (QUIJOTE)\citep{cernicharo2021_indene} and the GBT Observations of TMC-1: Hunting Aromatic Molecules (GOTHAM) \citep{mcguire2018}. The complete list of interstellar cyclic molecules includes various cyano (-CN) or ethynyl (-CCH) derivatives, where one of the hydrogen atoms is replaced by a CN or CCH radical, highlighting benzonitrile (\ch{c-C6H5CN}),\citep{mcguire2018} 1- and 2-cyanonaphthalene (\ch{c-C10H7CN}),\citep{mcguire2021} the five-membered ring 1-cyanocyclopentadiene (\ch{c-C5H5CN}),\citep{mccarthy2021} ethynyl-cyclopropenylidene (\ch{c-C3H-CCH}),\citep{cernicharo2021_indene}, two isomers of ethynyl-cyclopentadiene (\ch{c-C5H5CCH}),\citep{cernicharo2021_ethynylcyclopentadiene} and its structural isomer fulvenallene (\ch{c-C5H4CCH2}) \citep{cernicharo2022_fulvenallene}. Very recently, several isomers of the cyano-substituted derivatives of the three- and four-ring acenaphthylene (c-\ce{C12H7})\citep{Cernicharo2024} and pyrene ($c$-\ch{C16H9})\citep{Wenzel2024a,Wenzel2024b} have also been identified toward TMC-1, ranking among the largest molecules ever detected in space.\\ 
In this context, while unsaturated hydrocarbons with three, five, and six carbon atoms have been successfully detected in the ISM, the more simple but also highly strained cyclobutadiene (also known as [4]annulene, \ch{c-C4H4}) still remains uncharted. To date only its structural isomer vinylacetylene (\ch{CH2CHCCH}, global minimum in energy),\citep{Cremer2006} has been detected in the ISM, along with the related cyano derivative isomers \ch{H2CCHCCCN} and \textit{trans}-\ch{HCNCCHCCH}.\citep{Cernicharo2021c,lee2021b} This species is followed in energy by a cumulene form, butatriene (\ch{H2CCCCH2}) located at 7.7 kcal mol$^{-1}$ above \ch{CH2CHCCH} at the CCSD(T)/cc-pVTZ level, while \ch{c-C4H4} is found as the fourth most stable isomer, located at 33.4 kcal mol$^{-1}$.\citep{Cremer2006} Its inherent instability can be rationalized in terms of the presence of 4n delocalized $\pi$ electrons in it, which result in a so-called antiaromatic species. Nevertheless, in a recent study, \citealt{Wang2024} have presented the first bottom-up formation of \ch{c-C4H4} in low-temperature acetylene (\ch{C2H2}) ices exposed to energetic electrons, which simulates secondary electrons produced by the passage of cosmic rays. Once formed, \ch{c-C4H4} could perhaps be released into the gas phase through various thermal and non-thermal processes, serving as a promising candidate for astronomical searches, including James Webb Space Telescope (JWST) observations. Alternatively, its derivatization will infer a sizable dipole moment to the molecule and, therefore, its detection via rotational spectroscopy -both in the laboratory and in space- would be feasible. This, in turn, could serve as indirect evidence of the presence of \ch{c-C4H4} in the ISM.\\
Overall, the formation of CN and CCH derivatives of unsaturated hydrocarbons is suggested to proceed through the reaction of the hydrocarbon with the corresponding radicals, which occur rapidly at low temperatures \citep{Vakhtin2001}. This type of process has been proposed as pathways for the formation of \ch{c-C6H5CN},\citep{mcguire2018} \ch{c-C10H7CN},\citep{mcguire2021} and \ch{c-C5H5CN}, \citep{mccarthy2021} and are also thought to drive the relative abundance of the cyanopyrene isomers found in TMC-1. \citep{Wenzel2024b} In addition, a theoretical study explored the possibility of forming derivatives of cyclopropenylidene \ch{c-C3HX}. Theoretically, the reaction of \ch{c-C3H2} with 16 radicals was analyzed, and four cyclopropenylidene derivatives (\ch{c-C3HX}, \ch{X} = \ch{CN}, \ch{OH}, \ch{F}, \ch{NH2}) were proposed as potential species for detection. \citep{Flint2022} In a following study, these authors investigated the possible formation of cyclopropenylidene disubstituted with CN and CCH radicals in space.\citep{Flint2023}\\
%\citet{Wenzel2024b} concluded, based on theoretical calculations, that the relative abundances of cyanopyrene isomers are explained by the direct CN addition to pyrene under kinetic control in hydrogen-rich gas at 10 K. 
In this work, we have analyzed the formation processes of cyano, ethynyl and hydroxy derivatives of the structural isomers cyclobutadiene (\ch{c-C4H4}), butatriene(\ch{H2CCCCH2}), and vinylacetylene (\ch{H2CCHCCH}) through the following reaction: \ch{C4H4 + X → C4H3X + H}, where X = CN, CCH, and OH.  We highlight that all of these radicals have already been detected in the interstellar medium, further motivating the exploration of the aforementioned routes. We have also studied the thermochemistry of the different processes and analyzed the critical points of each potential energy surface (PES) to determine possible activation barriers.\\

\section{Computational methods}

\textit{Ab initio} and Density Functional Theory (DFT) methodologies have been employed to study the hydroxy, cyano, and ethynyl derivatives of cyclobutane (\ch{c-C4H4}), butatriene (\ch{H2CCCCH2}), and vinylacetylene (\ch{H2CCHCCH}), along with all the intermediates and transition states identified on the respective potential energy surfaces (PES) associated with their formation reactions. We note that the selection of computational levels prioritizes maximum accuracy while maintaining reasonable computational efficiency. \\
At the DFT level, we have chosen the hybrid functionals M08HX \citep{Zhao2008} and MPWB1K, \citep{Zhao2004} which, generally provide excellent results in both thermochemistry and kinetics. Additionally, the double-hybrid B2PLYPD3 functional \citep{Grimme2006} was employed. This functional includes Hartree-Fock exchange and a perturbative second-order correlation part, together with a Grimme’s D3BJ empirical dispersion term.\citep{Grimme2011} For these three functionals, Dunning’s correlation consistent triple-zeta basis sets, aug-cc-pVTZ, \citep{Dunning:2001dz} was used, which includes both polarization and diffuse functions on all elements.\\
For \textit{ab initio} calculations, we have selected the explicitly correlated coupled cluster theory with single and double excitations, including triplet excitations through a perturbative treatment, known as CCSD(T)-F12\citep{Knizia2009}, in conjunction with the cc-pVTZ-F12 basis set.\citep{Peterson2008} Recent studies \citep{Redondo2024} have shown that CCSD(T)-F12/cc-pVTZ-F12 optimized structures and energies are in excellent agreement with that obtained using a more complete “composite” scheme, but with a much lower computational cost. %This composite method starts from the CCSD(T)/cc-pVTZ results and incorporate corrections for basis set truncation error, diffuse function, and core-valence correlation. 
The calculated T1 diagnostic at the CCSD level \citep{Lee1989} was found to be below 0.02 for all characterized structures. This value is within the accepted threshold 
for systems that are well described by a single-reference wavefunction, supporting the reliability of our single-reference calculations.\\
Harmonic vibrational frequency calculations were carried out for each optimized geometries at every employed level of theory. This analysis allowed us to characterize the structures as either minima (all real frequencies) or transition states (with one imaginary frequency) and to determine the Zero Point Vibrational Energy (ZPVE). To verify that transition states connect the correct minima, intrinsic reaction coordinate (IRC) calculations \citep{Fukui1981} were performed for each transition state. Additionally, single-point energy calculations at the CCSD(T)-F12/cc-pVTZ-F12 level were performed on the DFT-optimized structures.\\
All calculations were performed using the GAUSSIAN 16,\citep{Frisch2016} MOLPRO \citep{werner2019} program packages, which already implement the needed methods, basis sets, and geometry optimization procedures.

\section{Results and discussion}

In this section, we will first analyze the geometrical parameters and the stabilities of the different isomers obtained by substituting a hydrogen atom with the hydroxyl (OH), cyano (CN), and ethynyl (CCH) radicals in cyclobutadiene, butatriene and vinylacetylene hydrocarbons. Afterward, we will examine the reactions between these hydrocarbons and the three radicals, which allows us to assess the feasibility of forming such substituted species.

To streamline the description of the computational levels employed here, we will adopt the following notation throughout the discussion: M08HX, MPWB1K, and B2PLYPD3 refer to calculations performed at these levels in conjunction with the aug-cc-pVTZ basis set; CCSD(T)-F12 denotes calculations performed at the CCSD(T)-F12/cc-pVTZ-F12 level; and CC-F12//M08HX, CC-F12//MPWB and CC-F12//B2PLD refer to energy calculations at the CCSD(T)-F12/cc-pVTZ-F12 level using geometries optimized at the M08HX/aug-cc-pVTZ, MPWB1K/aug-cc-pVTZ, and B2PLYPD3/aug-cc-pVTZ levels, respectively.

\subsection{Structure and stability of the OH, CN, and CCH derivatives of \ch{C4H4} isomers}

\begin{figure}[h]
\centering
  \includegraphics[height=8cm]{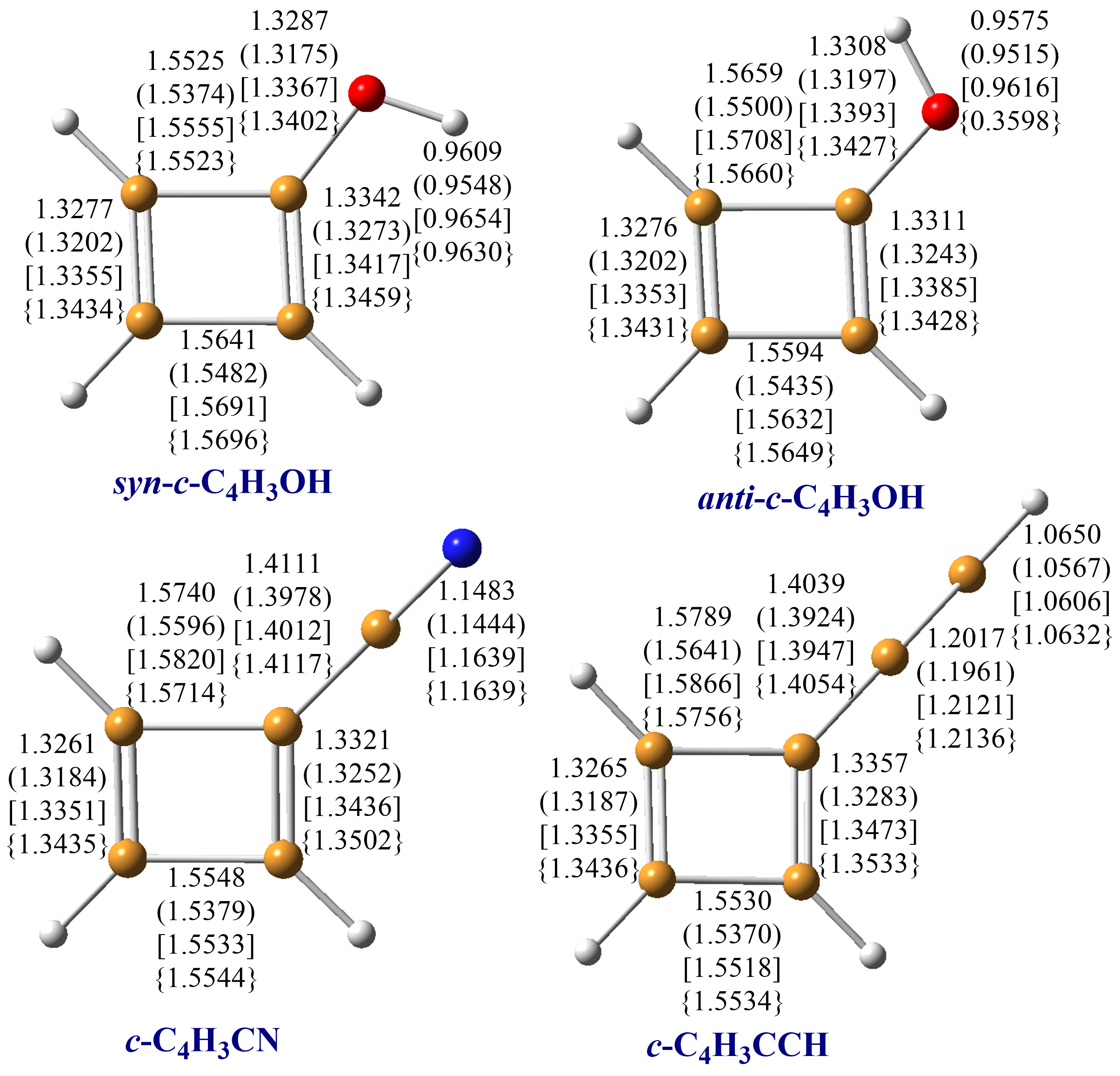}
  \caption{Geometrical parameters of the cyclobutadiene derivates, \ch{c-C4H3X} with X= OH, CN, and CCH, calculated at the M08HX/aug-cc-pVTZ, MPWB1K/aug-cc-pVTZ (in parentheses), B2PLYPD3/aug-cc-pVTZ (in brackets) and CCSD(T)-F12/cc-pVTZ-F12 (in curly bracket) levels. Distances are given in Angstroms.}
  \label{fgr:c-c4h3-X}
\end{figure}

\begin{figure}[h]
\centering
  \includegraphics[height=5.5cm]{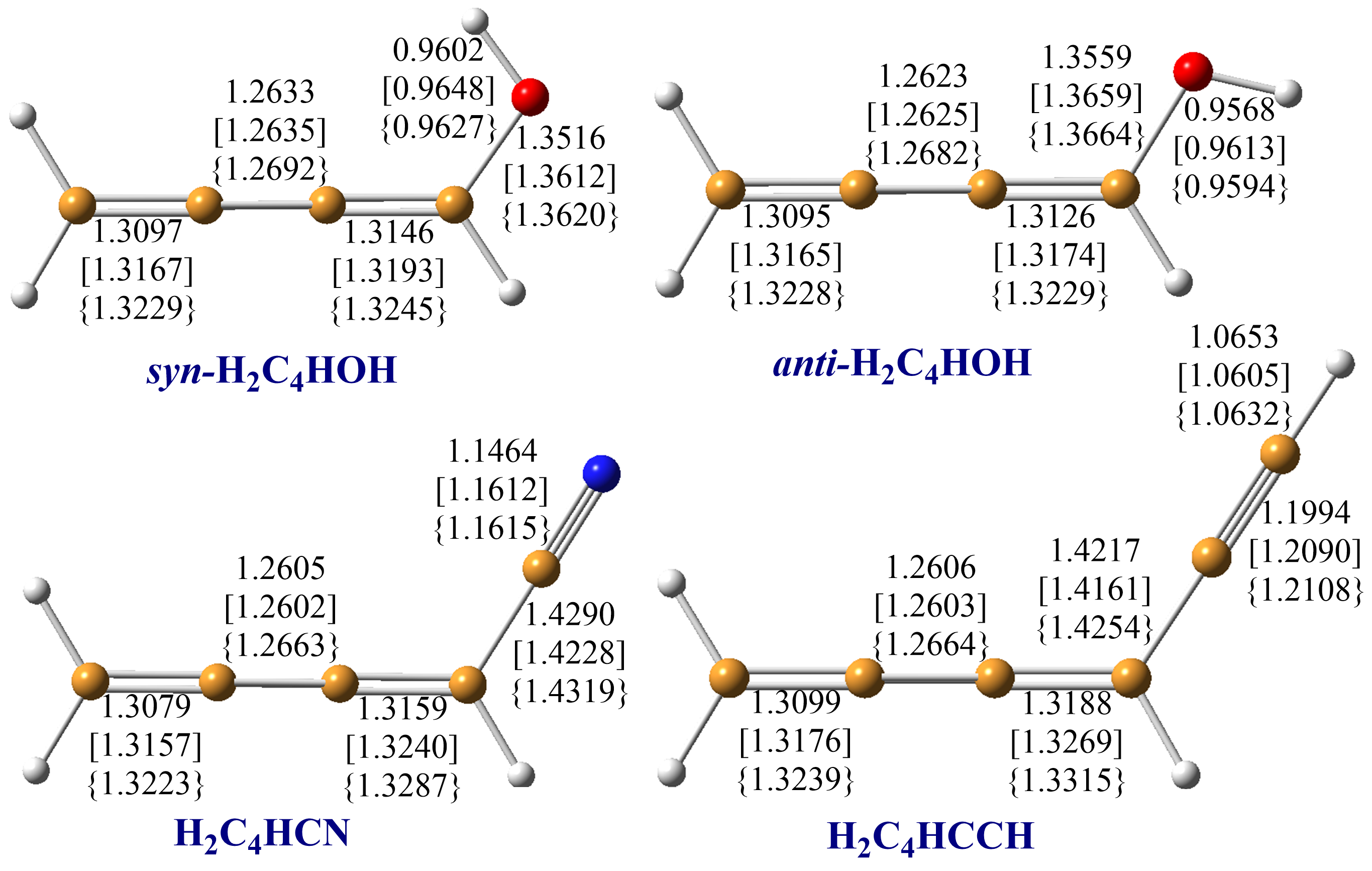}
  \caption{Geometrical parameters of the butatriene derivates, \ch{H2CCCCHX} with X= OH, CN, and CCH, calculated at the M08HX/aug-cc-pVTZ, B2PLYPD3/aug-cc-pVTZ (in brackets) and CCSD(T)-F12/cc-pVTZ-F12 (in curly bracket) levels. Distances are given in Angstroms.}
  \label{fgr:h2c4h-X}
\end{figure}

\begin{figure*}
 \centering
 \includegraphics[height=20cm]{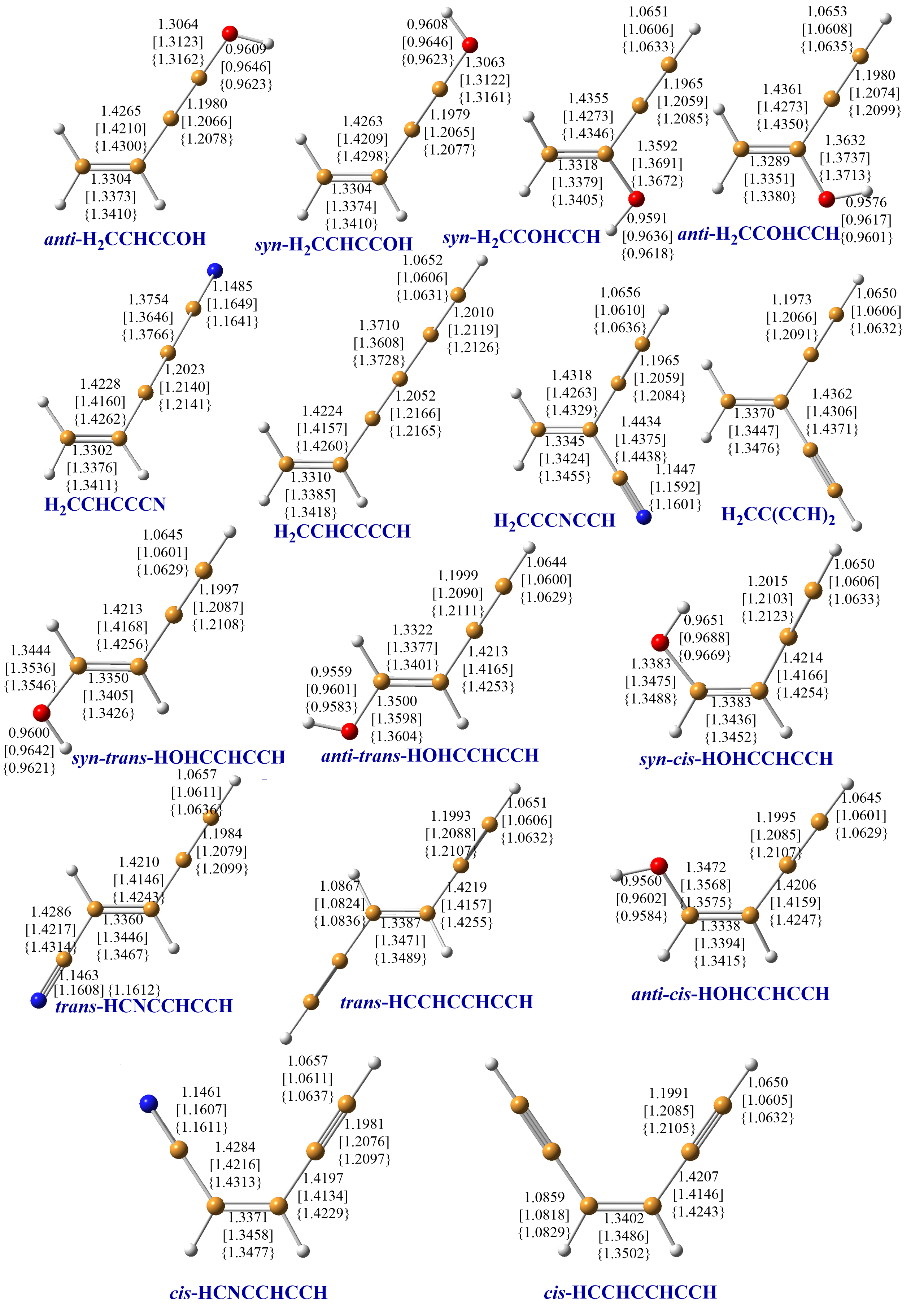}
 \caption{Geometrical parameters of the vinylacetylene derivates, \ch{H2CCHCCH(X)} with X= OH, CN, and CCH, calculated at the M08HX/aug-cc-pVTZ, B2PLYPD3/aug-cc-pVTZ (in brackets) and CCSD(T)-F12/cc-pVTZ-F12 (in curly bracket) levels. Distances are given in Angstroms.}
 \label{fgr:hccchch-X}
\end{figure*}

The geometries of the cyclobutadiene derivatives, 2-Hydroxycyclobuta-1,3-diene (\ch{c-C4H3OH}), 2-Cyanocyclobuta-1,3-diene (\ch{c-C4H3CN}), and 2-Ethynylcyclobuta-1,3-diene (\ch{c-C4H3CCH}) are shown in Fig.~\ref{fgr:c-c4h3-X}, while the structures of the 1-Hydroxybuta-1,2,3-triene (\ch{H2C4HOH}), 1-Cyanobuta-1,2,3-triene (\ch{H2C4HCN}), and 1-Ethynylbuta-1,2,3-triene (\ch{H2C4HCCH}) are given in Fig.~\ref{fgr:h2c4h-X}. Since the four hydrogen atoms of vinylacetylene are non-equivalent, the substitution of one of them will lead to four different structures. For instance, the cyano-substituted compounds identified are 1-vinylcyano-acetylene (\ch{H2CCHCCCN}), 1-vinyl-1-cyano-acetylene (\ch{H2CCCNCCH}), \textit{trans}-1-vinyl-2-cyano-acetylene (\textit{trans}-\ch{HCNCCHCCH}) and \textit{cis}-1-vinyl-2-cyano-acetylene (\textit{cis}-\ch{HCNCCHCCH}). The optimized structures of these vinylacetylene derivatives are shown in Fig.~\ref{fgr:hccchch-X}. For all the hydroxyl isomers, we have identified two conformers, distinguished by the orientation of the hydrogen atom in the hydroxyl group. These conformers are denoted with the prefixes \textit{syn} or \textit{anti}, indicating whether the hydrogen is oriented towards the double bond or in the opposite direction, respectively. The relative energies for the monosubstituted hydrocarbons obtained at the different computational levels are provided in Table~\ref{tbl:table1}. 

Additionally, as a reference, we report the optimized geometries of cyclobutadiene, butatriene, and vinylacetylene, as well as those of the OH, CN, and CCH radicals, at the levels of calculation employed in this study. These geometries are reported as Supplementary Information in Fig. S1, and the relative energies for the hydrocarbons are summarized in Table S1. The relative stability data given in Table S1 confirm that the most stable isomer is vinylacetylene, followed by butatriene and cyclobutadiene. The relative energy values calculated at the CC-F12//M08HX and CC-F12//B2PLD levels are nearly identical (7.64, 7.74 kcal mol$^{-1}$ for butatriene, and 33.22 and 33.06 kcal mol$^{-1}$ for cyclobutadiene, relative to vinylacetylene) and differ by less than 0.3 kcal mol$^{-1}$ from previously reported values at the CCSD(T) level using B3LYP-optimized geometries.\citep{Cremer2006,Wang2024} Our results obtained at the CCSD(T)-F12 level place butatriene at 9.03 kcal mol$^{-1}$ and cyclobutadiene at 32.97 kcal mol$^{-1}$ above vinylacetylene. This indicates that optimizing the geometries at the CCSD(T)-F12 level has a higher impact on the energy than changing the type of functional. This fact is consistent with the differences observed in the geometries at the various levels (Fig. S1), where it can be seen that the geometries optimized at the CCSD(T)-F12 level are the closest to the experimental data \citep{Hellwege1976,Huber1979,Kuchitsu1998}. The variations between CCSD(T)-F12 and experimental values for the C-C bond lengths in the \ch{C4H4} isomers are on the order of hundredths of an angstrom.

% referencias experimentales para incluirlas:
%C2CCCCH2 y H2CCHCCH: Hellwege, KH and AM Hellwege (ed.). Landolt-Bornstein: Group II: Atomic and Molecular Physics Volume 7: Structure Data of Free Polyatomic Molecules. Springer-Verlag. Berlin. 1976.
%CCH: K Kuchitsu(ed) "Structure of Free Polyatomic Molecules - Basic Data" Springer, Berlin, 1998
%OH and CN: Huber, K.P.; Herzberg, G., Molecular Spectra and Molecular Structure. IV. Constants of Diatomic Molecules, Van Nostrand Reinhold Co., 1979

\begin{table*}
\small
\tabcolsep 3.0pt
  \caption{ Relative energies (in kcal mol$^{-1}$) of the hydroxy, cyano and ethynyl derivatives of cyclobutadiene, butatriene and vinylacetylene computed at different levels. ZPV energies included}
  \label{tbl:table1}
  \begin{tabular*}{\textwidth}{@{\extracolsep{\fill}}lrrrrr}
    \hline
    Molecule & M08HX & CC-F12//M08HX & B2PLYPD3 & CC-F12//B2PLD & CCSD(T)-F12 \\
    \hline
    \textit{OH derivatives from the \ch{c-C4H4}/\ch{H2CCCCH2}/\ch{H2CCHCCH} + \ch{OH} reaction} & & & & & \\
    \textit{anti}-\ch{c-C4H3OH} ($^{1}$A') & 37.28 & 34.62 & 39.41 & 34.52 & 35.66\\
    \textit{syn}-\ch{c-C4H3OH} ($^{1}$A') & 36.46 & 33.81 & 38.55 & 33.70 & 34.55\\
    \textit{syn}-\ch{H2C4HOH} ($^{1}$A') & 10.89 & 12.29 & 10.50 & 12.40 & 13.41\\
    \textit{anti}-\ch{H2C4HOH} ($^{1}$A') & 13.19 & 14.54 & 12.88 & 14.68 & 16.18\\
    \textit{anti}-\ch{H2CCHCCOH} ($^{1}$A') & 12.68 & 13.52 & 13.39 & 13.67 & 14.67\\
    \textit{syn}-\ch{H2CCHCCOH} ($^{1}$A') & 12.67 & 13.52 & 13.37 & 13.66 & 14.70\\
    \textit{syn}-\ch{H2CCOHCCH} ($^{1}$A') & 5.86 & 4.81 & 5.26 & 4.73 & 6.02\\
    \textit{anti}-\ch{H2CCOHCCH} ($^{1}$A') & 4.89 & 3.95 & 4.45 & 3.97 & 5.39\\
    \textit{syn-trans}-\ch{HOHCCHCCH} ($^{1}$A') & 3.33 & 3.10 & 3.09 & 3.12 & 3.73\\
    \textit{anti-trans}-\ch{HOHCCHCCH} ($^{1}$A') & 3.97 & 3.75 & 3.78 & 3.77 & 4.77\\
    \textit{syn-cis}-\ch{HOHCCHCCH} ($^{1}$A') & 0.00 & 0.00 & 0.00 & 0.00 & 0.00 \\
    \textit{anti-cis}-\ch{HOHCCHCCH} ($^{1}$A') & 3.89 & 3.52 & 4.12 & 4.13 & 4.68\\
    \textit{CN derivatives from the \ch{c-C4H4}/\ch{H2CCCCH2}/\ch{H2CCHCCH} + \ch{CN} reaction} & & & & & \\
    \ch{c-C4H3CN} ($^{1}$A') & 33.56 & 30.36 & 35.21 & 30.20 & 31.14\\
    \ch{H2C4HCN} ($^{1}$A') & 7.08 & 9.03 & 7.65 & 9.16 & 9.05\\
    \ch{H2CCHCCCN} ($^{1}$A') & -0.23 & 0.05 & -1.48 & 0.13 & 0.07\\
    \ch{H2CCCNCCH} ($^{1}$A') & 3.81 & 3.00 & 3.58 & 2.91 & 2.92\\
    \textit{trans}-\ch{HCNCCHCCH} ($^{1}$A$_{g}$) & 0.00 & 0.00 & 0.00 & 0.00 & 0.00\\
    \textit{cis}-\ch{HCNCCHCCH} ($^{1}$A$_{g}$) & 0.45 & 0.34 & 0.28 & 0.34 & 0.32\\
    \textit{CCH derivatives from the \ch{c-C4H4}/\ch{H2CCCCH2}/\ch{H2CCHCCH} + \ch{CCH} reaction} & & & & & \\ 
    \ch{c-C4H3CCH} ($^{1}$A') & 36.11 & 32.36 & 38.67 & 32.12 & 32.52\\
    \ch{H2C4HCCH} ($^{1}$A') & 10.13 & 11.46 & 11.64 & 11.56 & 11.51\\
    \ch{H2CCHCCCCH} ($^{1}$A') & 0.00 & 0.00 & 0.00 & 0.00 & 0.00\\
    \ch{H2CC(CCH)2} ($^{1}$A') & 6.67 & 5.16 & 7.47 & 5.07 & 5.19\\
    \textit{trans}-\ch{HCCHCCHCCCH} ($^{1}$A$_{g}$) & 2.89 & 2.33 & 3.94 & 2.28 & 2.34\\
    \textit{cis}-\ch{HCCHCCHCCCH} ($^{1}$A$_{g}$) & 3.22 & 2.58 & 4.17 & 2.61 & 2.63\\
    \hline
  \end{tabular*}
\end{table*}

We find that for monosubstituted hydrocarbons with OH, CN and CCH radicals, there are only slight variations in the structural parameters compared to the parent hydrocarbons. These changes are primarily observed in the carbon bond where the substitution occurs. Specifically, the C-C distance (or distances) adjacent to the substituted hydrogen increases, as shown in Fig.~\ref{fgr:c-c4h3-X},~\ref{fgr:h2c4h-X}, and ~\ref{fgr:hccchch-X}. Regarding the different calculation levels employed, it can be seen that the geometries optimized using the double hybrid functional B2PLYPD3 are the closest to those calculated at the CCSD(T)-F12 level, while the geometries obtained using the hybrid functional MPWB1K are the furthest. Therefore, the latter has only been used for cyclobutadiene derivatives.

The relative stability of the isomers \ch{c-C4H3X}, \ch{H2C4HX}, and \ch{H2CCHCC(X)}, with X = OH, CN and CCH (Table~\ref{tbl:table1}), follows a similar trend for all the radicals: the monosubstituted cyclobutadiene isomers are the most unstable species in all cases, while the monosubstituted vinylacetylene isomers are the most stable. This trend mirrors that observed for \ch{C4H4} hydrocarbons.

Afterward, we will focus on the stability of the four vinyl acetylene derivatives as function of the substitution position. The monosubstituted hydrocarbons, in one of the two hydrogen atoms attached to the terminal carbon of the vinyl group, \textit{cis}-\ch{HXCCHCCH} and \textit{trans}-\ch{HXCCHCCH} exhibit similar stabilities, with energy differences of less than 1 kcal mol$^{-1}$ for the CCH radical.  In contrast, substitution at the non-terminal hydrogen of the vinyl group leads to the most unstable compounds, \ch{H2CCXCCH}. For the hydroxy radical, the most favorable structures are obtained by substitution of the terminal hydrogens of the vinyl group, with the most stable isomer being \textit{syn-cis}-\ch{HOHCCHCCH}. When the ethynyl radical is involved, the substitution of the acetylenic hydrogen gives rise to the most stable structure, \ch{H2CCHCCCCH}. Whereas for the cyano radical, substitutions at the terminal hydrogens of the vinyl and acetylene groups results in quasi-isoenergetic isomers (the energy differences between \ch{H2CCHCCCN}, \textit{cis}-\ch{HCNCCHCCH} and \textit{trans}-\ch{HCNCCHCCH} isomers is less than 0.5 kcal mol$^{-1}$). These relative stabilities arise from various factors. The formation of \ch{H2CCHCCX} isomers where the radicals have $\pi$-type electrons favors delocalization along the carbon chain.  Conversely, more electrophilic substituents prefer to attack positions where the carbon has a larger negative partial charge, (i.e., the terminal carbon of the vinyl group). Another factor to take into consideration is the steric effect, which may makes substitution at less crowded positions more favorable when replacing a hydrogen atom with radicals.

An analysis of the relative energies at the different computational levels presented in Table~\ref{tbl:table1} shows that the CC-F12//M08HX and CC-F12//B2PLD results are closer to those obtained at the CCSD(T)-F12 level than those derived at the DFT levels (M08HX or B2PLYPD3). The results that most closely match the CCSD(T)-F12 data are those calculated using the B2PLYPD3 geometries (CC-F12//B2PLD), with energy differences of less than 0.5 kcal mol$^{-1}$ for the CN and CCH derivatives, and approximately 1 kcal mol$^{-1}$ for the OH derivatives.

\subsection{Formation of the OH, CN, and CCH derivatives of \ch{C4H4} isomers}

As a potential route for the formation of the hydroxyl, cyano and ethynyl derivatives of cyclobutadiene, butatriene and vinylacetylene, we have considered the following reaction: 

\small
\begin{equation}
  \ch{C4H4} + \ch{X} \rightarrow \ch{C4H3X} + \ch{H}
\end{equation}
\normalsize

where \ch{C4H4} represents the three isomers obeying to the general formula \ch{C4H4} and X denotes the OH, CN or CCH radicals.

The intermediates (I) and transition states (TS) located on the different potential energy surfaces (PES) are labeled using a subscript indicating the related hydrocarbon (\textit{cyc}, \textit{but}, or \textit{vin}, for reactions with cyclobutadiene, butatriene, or vinylacetylene) and the radical (OH, CN, or CCH) involved in the reaction under study. The reactions studied all proceed on the doublet PES. Geometric optimizations have been performed at the CCSD(T)-F12 level exclusively for reactants and products. The relative energies calculated at this level on the DFT-optimized geometries (CC-F12//DFT) are in close agreement with those obtained directly at the CCSD(T)-F12 level. In particular, the reaction energies computed at the CC-F12//B2PLD level generally differ by about 0.5 kcal mol$^{-1}$ from the CCSD(T)-F12 values.

\subsubsection{Reaction of cyclobutadiene with OH, CN, and CCH.}

\begin{figure*}
 \centering
 \includegraphics[height=11cm]{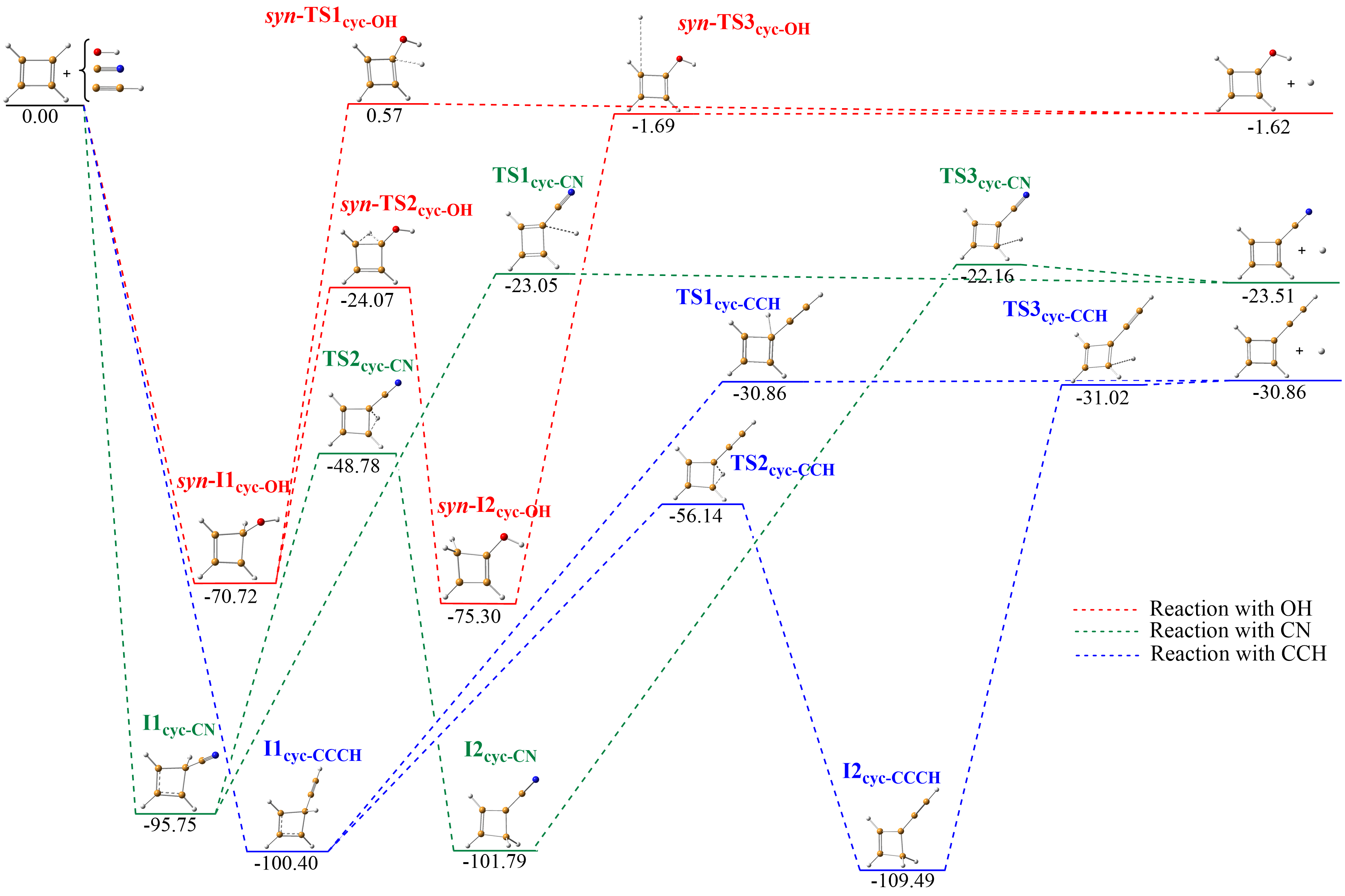}
 \caption{Relative energies (in kcal mol$^{-1}$) for the stationary points located along the gas phase reaction of cyclobutadiene with hydroxyl, cyano, and ethynyl radicals computed at the CCSD(T)-F12/cc-pVTZ-F12//B2PLYPD3/aug-cc-pVTZ level. ZPV energies included.}
 \label{fgr:c-c4h4+x}
\end{figure*}

The relative energies of the characterized stationary points for the reactions between cyclobutadiene and the OH, CN, and CCH radicals are shown in Table~\ref{tbl:table2} for the different calculation levels. The corresponding reaction profiles are schematically depicted in Fig.~\ref{fgr:c-c4h4+x}.

The formation of hydroxy, cyano, and ethynyl derivatives of cyclobutadiene are exothermic processes. The most favorable one is the production of \ch{c-C4H3CCH} followed by that of \ch{c-C4H3CN}, whose reaction energies are -30.27 and -22.62 kcal mol$^{-1}$ at CCSD(T)-F12 level, respectively. The formation of the two hydroxyl-substituted conformers is slightly exothermic (with energies of -0.58 and -1.51 kcal mol$^{-1}$ at the CCSD(T)-F12 level for \textit{anti}-\ch{c-C4H3OH} and \textit{syn}-\ch{c-C4H3OH}, respectively). Thus, from an energetic point of view, the formation would be possible at the low temperature conditions of the ISM, provided that barrierless reaction pathways are available.

The reaction mechanism for the substitution process is analogous for all three radicals, as illustrated in Fig.~\ref{fgr:c-c4h4+x}, which depicts the characterized stationary points in their respective potential energy surfaces (PESs). For simplicity, in the reaction involving the OH radical, only the reaction pathways corresponding to the \textit{syn} conformers have been represented, while the energy values for the \textit{anti} conformers are provided in Table~\ref{tbl:table2}.

As a representative example, we consider the reaction mechanism of cyclobutadiene with the CN radical. The process begins by the insertion of the CN radical into one of the carbons giving rise to the intermediate denoted as I1$_{cyc-CN}$ located -95.75 kcal mol$^{-1}$ below the reactants at the CC-F12//B2PLD level. This step does not involve any transition state. Once the intermediate I1$_{cyc-CN}$ is formed, the elimination of the hydrogen atom through transition state TS1$_{cyc-CN}$ (located about -23.05 kcal mol$^{-1}$ at the CC-F12//B2PLD level) leads to the product \ch{c-C4H3CN}.  Alternatively, the intermediate I1$_{cyc-CN}$ can undergo isomerization to form the more stable intermediate I2$_{cyc-CN}$ (-101.79 kcal mol$^{-1}$ at the CC-F12//B2PLD level) through hydrogen migration from the carbon where the CN insertion occurred to the adjacent carbon.This step involves the transition state TS2$_{cyc-CN}$ located -48.78 kcal mol$^{-1}$ below the reactants at the CC-F12//B2PLD level.  Subsequently, the elimination of one of the hydrogen atoms bonded to the carbon in I2$_{cyc-CN}$ also yields the formation of \ch{c-C4H3CN}. This step involves the transition state TS3$_{cyc-CN}$, which is clearly below the reactants (-22.16 kcal mol$^{-1}$ at the CC-F12//B2PLD level). Therefore, the formation of \ch{c-C4H3CN} via both pathways proceeds without a net activation barrier, indicating that the reaction is feasible under ISM conditions. The reaction mechanisms can be summarized as follows:

\small
\begin{equation}
\ch{c-C4H4} + \ch{CN} \rightarrow \mathrm{I1_{cyc-CN}} \rightarrow \mathrm{TS1_{cyc-CN}} \rightarrow \ch{c-C4H3CN} + \ch{H}
\end{equation}
\normalsize

\small
\begin{equation}
\begin{split}
\ch{c-C4H4} + \ch{CN} \rightarrow \mathrm{I1_{cyc-CN}} \rightarrow \mathrm{TS2_{cyc-CN}}  \rightarrow  \mathrm{I2}_{cyc-CN} \\
\rightarrow \mathrm{TS3}_{cyc-CN} \rightarrow \ch{c-C4H3CN} + \ch{H}
\end{split}
\end{equation}
\normalsize

Regarding the reaction between cyclobutadiene and the ethynyl radical, it shows a reaction profile analogous to that observed for the corresponding reaction with the cyano radical. However, in the case of the CCH radical, the identified stationary points and the final product possess greater relative stability compared to those found in the CN system, with energy differences ranging from 10 to 20 kcal mol$^{-1}$. As illustrated in Fig.~\ref{fgr:c-c4h4+x}, the transition states TS1$_{cyc-CCH}$ and TS3$_{cyc-CCH}$, when zero-point vibrational energy (ZPVE) corrections are included, show energies equal to or slightly lower than that of the reaction product, \ch{c-C4H3CCH}.

As previously indicated, the formation of the two conformers of the hydroxy derivatives, \textit{syn}- and \textit{anti}-\ch{c-C4H3OH}, is only barely exothermic. As shown in Table~\ref{tbl:table2} and Fig.~\ref{fgr:c-c4h4+x}, both \textit{syn}- and \textit{anti}-TS1$_{cyc-OH}$ transition states lie slightly above the reactants (0.57 kcal mol$^{-1}$ and 1.29 kcal mol$^{-1}$ at the CC-F12//B2PLD, respectively), suggesting that the formation of the hydroxy derivative could proceed in the ISM via the \textit{syn}- and \textit{anti}-I2$_{cyc-OH}$ intermediates. This path involves the transition states \textit{syn}- and \textit{anti}-TS3$_{cyc-OH}$ lying below the reactants (-0.83 kcal mol$^{-1}$ and -1.69 kcal mol$^{-1}$ at the CC-F12//B2PLD, respectively). In Table~\ref{tbl:table2}, for the reaction of \ch{c-C4H4} with radical OH we include transition states TS4-I1$_{cyc-OH}$ and TS4-I2$_{cyc-OH}$, which correspond to the interconversion between the \textit{syn} and \textit{anti} conformers of intermediates I1$_{cyc-OH}$ and I2$_{cyc-OH}$, respectively. As shown, the activation barriers for interconversion are significantly lower than the energy of the reactants (-70.13 and -72.60 kcal mol$^{-1}$ at the CC-F12//B2PLD level of theory, respectively). This suggests that interconversion to the more stable conformer is energetically favorable under ISM conditions.

In summary, the results for the reaction of cyclobutadiene with OH, CN, and CCH radicals indicate that the formation of cyano, ethynyl, and hydroxy derivatives of cyclobutadiene is  thermodynamically feasible and involves no net activation barriers. Among these, the most favorable processes involve the formation of \ch{c-C4H3CN} and \ch{c-C4H3CCH} derivatives, which appears as thrilling candidates for interstellar detection. Moreover, this derivatization infers a significant dipole moment to the parental and apolar \ch{c-C4H4}, enabling its detection by means of rotational spectroscopy in the laboratory but also in space. In this context, the main challenge for such measurements lies in generating sufficient quantities of these highly unstable antiaromatic species in the gas phase. However, emerging laboratory techniques, such as direct-absorption millimeter and submillimeter spectroscopy of desorbed species from ice samples,\citep{Milam2020} may succeed where conventional spectroscopic methods are likely to fail. Furthermore, the eventual detection of CN and CCH derivatives could serve as indirect evidence for the presence of \ch{c-C4H4} in the interstellar medium (ISM).

\begin{table*}
\small
  \caption{Relative energies (in kcal mol$^{-1}$) for the stationary points located along the gas-phase reaction paths of cyclobutadiene with hydroxyl, cyano, and ethynyl radicals computed at different levels. ZPV energies included}
  \label{tbl:table2}
  \begin{tabular*}{\textwidth}{@{\extracolsep{\fill}}lrrrrrrr}
    \hline
    Molecule & M08HX & CC-F12//M08HX & MPWB1K & CC-F12//MPWB & B2PLYPD3 & CC-F12//B2PLD & CCSD(T)-F12 \\
    \hline
    \textit{Reaction \ch{c-C4H4} + \ch{OH}} & & & & & & & \\
    \ch{c-C4H4} + \ch{OH} & 0.00 & 0.00 & 0.00 & 0.00 & 0.00 & 0.00 & 0.00 \\
    \textit{anti}-\ch{c-C4H3OH} + \ch{H} & -2.86 & -0.66 & -2.44 & -0.57 & -1.17 & -0.80 & -0.58 \\
    \textit{syn}-\ch{c-C4H3OH} + H & -3.68 & -1.48 & -3.37 & -1.42 & -2.03 & -1.62 & -1.51 \\
    \textit{anti}-I1$_{cyc-OH}$ & -75.96 & -73.02 & -78.08 & -72.93 & -74.05 & -72.51	& \\
    \textit{syn}-I1$_{cyc-OH}$ & -74.32 & -71.26 & -76.35 & -71.14 & -72.30 & -70.72 & \\
    \textit{anti}-I2$_{cyc-OH}$ & -80.75 & -75.78 & -83.09 & -75.78 & -77.88 & -75.42 & \\
    \textit{syn}-I2$_{cyc-OH}$ & -80.51 & -75.61 & -82.92 & -75.64 & -77.71 & -75.30 & \\
    \textit{anti}-TS1$_{cyc-OH}$ & -1.78 & 0.02 & -2.01 & -0.23 & 1.94 & 1.29 & \\
    \textit{syn}-TS1$_{cyc-OH}$ & -2.27 & -0.50 & -3.13 & -0.61 & 0.79 & 0.57 & \\
    \textit{anti}-TS2$_{cyc-OH}$ & -28.51 & -22.95 & -30.10 & -22.99 & -25.81 & -22.95 & \\
    \textit{syn}-TS2$_{cyc-OH}$ & -29.73 & -24.20 & -31.29 & -24.13 & -26.99 & -24.07 & \\
    \textit{anti}-TS3$_{cyc-OH}$ & -3.36 & -0.56 & -2.48 & -1.01 & -1.24 & -0.83 & \\
    \textit{syn}-TS3$_{cyc-OH}$ & -4.03 & -1.33 & -3.31 & -1.75 & -2.07 & -1.69 & \\  
    TS4-I1$_{cyc-OH}$ & -73.76 & -70.58 & -75.79 & -70.56 & -71.68 & -70.13 & \\
    TS5-I2$_{cyc-OH}$ & -77.63 & -72.98 & -79.93 & -73.01 & -74.69 & -72.60 & \\
    \textit{Reaction \ch{c-C4H4} + \ch{CN}} & & & & & & & \\
    \ch{c-C4H4} + \ch{CN} & 0.00 & 0.00 & 0.00 & 0.00 & 0.00 & 0.00 & 0.00\\
    \ch{c-C4H3CN} + \ch{H} & -30.88 & -23.65 & -31.40 & -23.64 & -28.07 & -23.51 & -22.62 \\
    I1$_{cyc-CN}$ & -106.45 & -96.48 & -109.45 & -96.47 & -101.40 & -95.75 & \\
    I2$_{cyc-CN}$ & -113.45 & -102.76 & -117.48 & -102.75 & -108.48 & -101.79 & \\
    TS1$_{cyc-CN}$ & -30.26 & -22.74 & -34.06 & -22.43 & -23.82 & -23.05 & \\
    TS2$_{cyc-CN}$ & -61.07 & -49.36 & -64.01 & -49.34 & -55.94 & -48.78 & \\
    TS3$_{cyc-CN}$ & -31.20 & -23.58 & -31.06 & -23.55 & -23.64 & -22.16 & \\
    \textit{Reaction \ch{c-C4H4} + \ch{CCH}} & & & & & & & \\ 
    \ch{c-C4H4} + \ch{CCH} & 0.00 & 0.00 & 0.00 & 0.00 & 0.00 & 0.00 & 0.00 \\
    \ch{c-C4H3CCH} + \ch{H} & -34.39 & -30.90 & -35.66 & -31.01 & -35.60 & -30.86 & -30.27 \\
    I1$_{cyc-CCH}$ & -107.08 & -101.08 & -110.74 & -101.16 & -106.00 & -100.40 & \\
    I2$_{cyc-CCH}$ & -117.35 & -110.08 & -122.10 & -110.52 & -116.28 & -109.49 & \\
    TS1$_{cyc-CCH}$ & -33.73 & -30.07 & -39.12 & -57.26 & -32.77 & -30.86 & \\
    TS2$_{cyc-CCH}$ & -64.67 & -56.66 & -68.25 & -56.75 & -63.38 & -56.14 & \\    
    TS3$_{cyc-CCH}$ & -34.85 & -30.94 & -35.56 & -31.16 & -35.86 & -31.02 & \\
    \hline
  \end{tabular*}
\end{table*}

\subsubsection{Reaction of butatriene with OH, CN, and CCH.}

\begin{figure*}
 \centering
 \includegraphics[height=11cm]{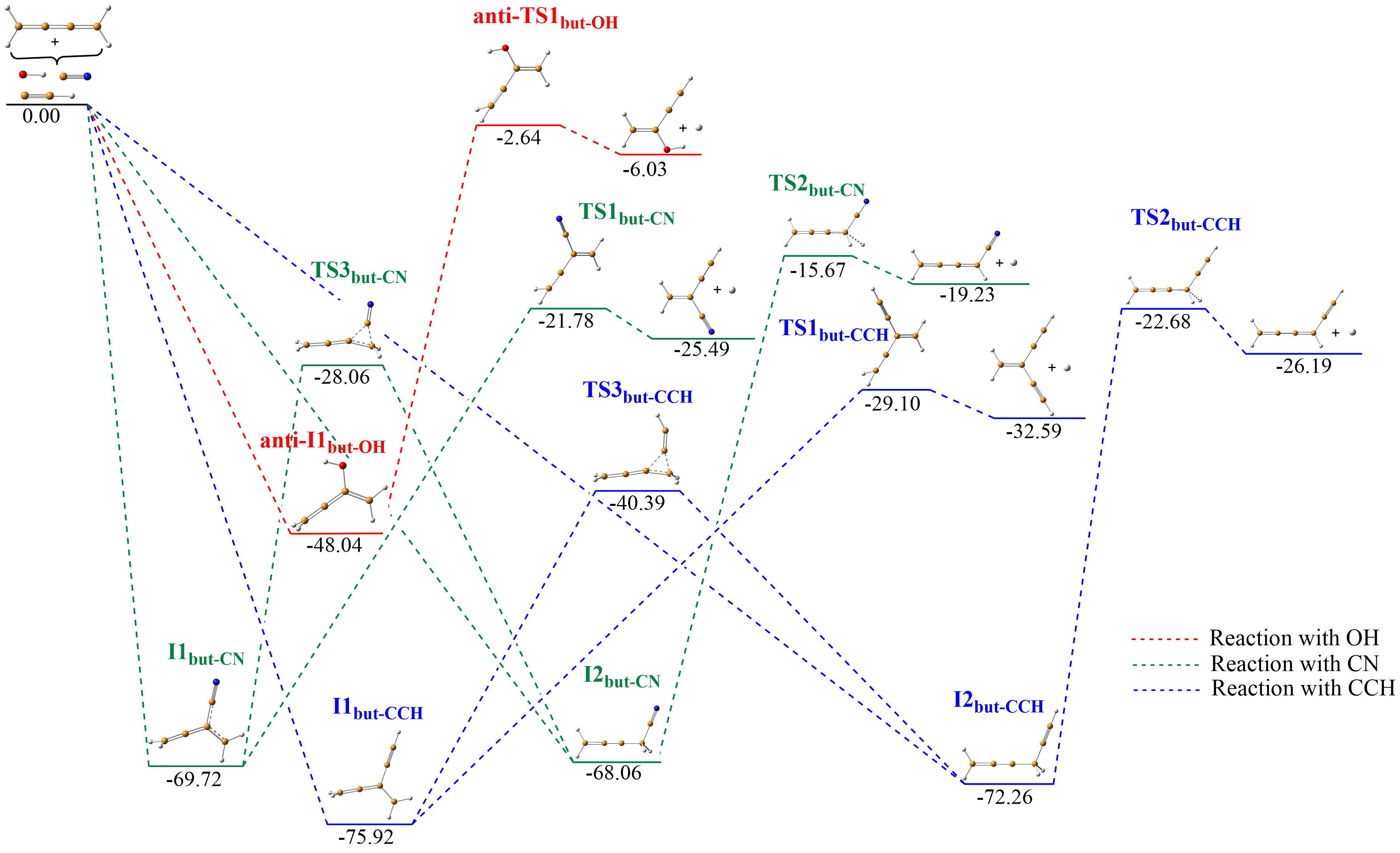}
 \caption{Relative energies (in kcal mol$^{-1}$) for the stationary points located along the gas phase reaction of butatriene with cyano, and ethynyl radicals computed at the CCSD(T)-F12/cc-pVTZ-F12//B2PLYPD3/aug-cc-pVTZ level. ZPV energies included.}
 \label{fgr:h2c4h2+X}
\end{figure*}

The stationary points on the doublet potential energy surfaces (PESs) corresponding to the reactions between butatriene and the hydroxy, cyano, and ethynyl radicals at various computational levels are summarized in Table~\ref{tbl:table3}. The reaction profiles are depicted in Fig.~\ref{fgr:h2c4h2+X}. As in the case of cyclobutadiene reacting with OH, only the pathway associated with the \textit{anti} conformer is presented, as it is the most thermodynamically favorable.

As shown in Table~\ref{tbl:table3}, the formation of 1-cyanobuta-1,2,3-triene (\ch{H2C4HCN}) and 1-ethynylbuta-1,2,3-triene (\ch{H2C4HCCH}) are significantly exothermic, with reaction energies of -19.27 kcal mol$^{-1}$ and -25.85 kcal mol$^{-1}$, respectively, at the CCSD(T)-F12 level. On the other hand, the formation of 1-hydroxybuta-1,2,3-triene (\ch{H2C4HOH}) is slightly endothermic for both the \textit{syn} and \textit{anti} conformers, with computed reaction energies of 2.49 kcal mol$^{-1}$ and 4.75 kcal mol$^{-1}$, respectively, at the same level of theory. Consequently, the formation of these hydroxy-containing species is not feasible under the low-temperature conditions characteristic of the ISM. Additionally, we consider the vinyl acetylene derivatives 1-vinyl-1-cyano-acetylene (\ch{H2CCCNCCH}), 1-vinyl-1-ethynyl-acetylene (\ch{H2CC(CCH)2}), and the \textit{syn}- and \textit{anti}-1-vinyl-1-hydroxy-acetylene (\ch{H2CCOHCCH}) as potential reaction products, as they can also form through the interaction of butatriene with the aforementioned radicals. These reaction pathways are exothermic, with reaction energies of -25.40, -32.18, -5.06, and -5.86 kcal mol$^{-1}$, respectively, at the CCSD(T)-F12 level. The corresponding reaction pathways for their formation are also illustrated in Fig.~\ref{fgr:h2c4h2+X}.

We then analyze the reaction profile starting with the interaction between butatriene and the CN radical as a reference. The CN radical can either bond to one of the central carbon atoms of butatriene, leading to the intermediate I1$_{but-CN}$, or to one of the terminal carbon atoms, yielding intermediate I2$_{but-CN}$. An analysis of the reaction coordinate in both cases indicates that these intermediates form without passing through a transition state. Both are significantly more stable than the reactants, with relative energies of -69.72 kcal mol$^{-1}$ and -68.06 kcal mol$^{-1}$, respectively, at the CC-F12//B2PLD level. From I1$_{but-CN}$, the elimination of a hydrogen atom from the terminal carbon farthest from the CN group leads to the formation of the most exothermic product, the vinylacetylene derivative \ch{H2CCCNCCH}. This transformation occurs through the TS1$_{but-CN}$ transition state, which is located below the energy of the reactants (-21.78 kcal $^{-1}$ at the CC-F12//B2PLD level). The overall reaction process can be summarized as follows:

\small
\begin{equation}
\ch{H2C4H2} + \ch{CN} \rightarrow \mathrm{I1_{but-CN}} \rightarrow \mathrm{TS1_{but-CN}} \rightarrow \ch{H2CCCNCCH} + \ch{H}
\end{equation}
\normalsize

Alternatively, if one of the hydrogens in I1$_{but-CN}$ is removed from the carbon closest to the CN group, a cyclic product is formed. This process involves an activation barrier of more than 30 kcal mol$^{-1}$. For this reason, it has not been included, as it is, in principle, an unfeasible process under ISM conditions.

Starting from the intermediate I2$_{but-CN}$, the elimination of one hydrogen from the carbon to which the CN radical is attached leads to the formation of the butatriene derivative, \ch{H2C4HCN}. This process takes place through the TS2$_{but-CN}$ transition state, located -15.67 kcal mol$^{-1}$ below the reactants. The overall process can be summarized as:

\small
\begin{equation}
\ch{H2C4H2} + \ch{CN} \rightarrow \mathrm{I2_{but-CN}} \rightarrow \mathrm{TS2_{but-CN}} \rightarrow \ch{H2C4HCN} + \ch{H}
\end{equation}
\normalsize

As can be seen in Fig.~\ref{fgr:h2c4h2+X}, the intermediate I2$_{but-CN}$ can also isomerize to the more stable one, I1$_{but-CN}$, through the TS3$_{but-CN}$ transition state, which is located 28.06 kcal mol$^{-1}$ below the reactants.

Concerning the reaction of butatriene with the CCH radical, its profile closely parallels that of the reaction with the CN radical and, similarly to the reactions with cyclobutadiene, the various PES stationary points along the reaction paths are relatively more stable. Overall, the results derived for the reaction of butatriene with CN and CCH radicals indicate that the butatriene derivatives, \ch{H2C4HCN} and \ch{H2C4HCCH}, can be readily produced, but the formation of vinylacetylene derivatives, \ch{H2CCCNCCH} and \ch{H2CC(CCH)2}, is even more favorable from the energetic (more exothermic) point of view and shows lower barriers.

A similar results is obtained for the reaction of \ch{H2C4H2} with the OH radical, for which, from a thermodynamic point of view, only the formation of the vinylacetylene derivatives, \textit{syn}- and \textit{anti}-\ch{H2CCOHCCH}, is feasible (see Table~\ref{tbl:table3}). As shown in the Fig.~\ref{fgr:h2c4h2+X}, these pathways proceed via the \textit{syn}- and \textit{anti}-syn-I1$_{but-OH}$ intermediates by elimination of a hydrogen from the terminal carbon farthest from the OH group. The transition states involved, \textit{syn}- and \textit{anti}-TS1$_{but-OH}$, are located 1.86 and 2.64 kcal mol$^{-1}$ below the reactants. Therefore, the formation of  \textit{syn}- and \textit{anti}-\ch{H2CCOHCCH} is also feasible.

\begin{table*}
\small
  \caption{Relative energies (in kcal mol$^{-1}$) for the stationary points located along the gas-phase reaction paths of butatriene with hydroxyl, cyano, and ethynyl radicals computed at different levels. ZPV energies included}
  \label{tbl:table3}
  \begin{tabular*}{\textwidth}{@{\extracolsep{\fill}}lrrrrr}
    \hline
    Molecule & M08HX & CC-F12//M08HX & B2PLYPD3 & CC-F12//B2PLD & CCSD(T)-F12 \\
    \hline
    \textit{Reaction \ch{H2CCCCH2} + \ch{OH}} & & & & & \\
    \ch{H2C4H4} + \ch{OH} & 0.00 & 0.00 & 0.00 & 0.00 & 0.00 \\
    \textit{syn}-\ch{H2C4HOH} + \ch{H} & 1.60 & 2.59 & 2.41 & 2.40 & 2.49 \\
    \textit{anti}-\ch{H2C4HOH} + H & 3.90 & 4.83 & 4.80 & 4.68 & 4.75 \\
    \textit{syn}-\ch{H2CCOHCCH} + \ch{H} & -3.43 & -4.90 & -2.82 & -5.27 & -5.06\\
    \textit{anti}-\ch{H2CCOHCCH} + \ch{H} & -4.39 & -5.76 & -3.63 & -6.03 & -5.86 \\
    \textit{syn}-I1$_{but-OH}$ & -47.05 & -47.25 & -44.96 & -47.03 & \\
    \textit{anti}-I1$_{but-OH}$ & -48.19 & -48.33 & -46.15 & -48.04 & \\
    \textit{syn}-TS1$_{but-OH}$ & -0.26 & -1.97 & 0.05 & -1.89 & \\
    \textit{anti}-TS1$_{but-OH}$ & -1.11 & -2.77 & -0.71 & -2.64 & \\
    \textit{Reaction \ch{H2CCCCH2} + \ch{CN}} & & & & & \\
    \ch{H2C4H2} + \ch{CN} & 0.00 & 0.00 & 0.00 & 0.00 & 0.00 \\
    \ch{H2C4HCN} + \ch{H} & -26.50 & -19.40 & -23.13 & -19.23 & -19.27\\
    \ch{H2CCCNCCH} + \ch{H} & -29.78 & -25.44 & -27.20 & -25.49 & -25.40\\
    I1$_{but-CN}$ & -77.08 & -70.21 & -71.98 & -69.72 & \\
    I2$_{but-CN}$ & -75.77 & -68.43 & -70.50 & -68.06 & \\
    TS1$_{but-CN}$ & -26.54 & -22.33 & -23.73 & -21.78 & \\
    TS2$_{but-CN}$ & -23.07 & -16.37 & -18.61 & -15.67 & \\
    TS3$_{but-CN}$ & -38.48 & -28.90 & -30.84 & -28.06 & \\
    \textit{Reaction \ch{H2CCCCH2} + \ch{CCH}} & & & & & \\ 
    \ch{H2C4H2} + \ch{CCH} & 0.00 & 0.00 & 0.00 & 0.00 & 0.00 \\
    \ch{H2C4HCCH} + \ch{H} & -29.52 & -26.23 & -30.12 & -26.10 & -25.85 \\
    \ch{H2CC(CCH)2} + \ch{H} & -32.98 & -32.53 & -34.30 & -32.59 & -32.18\\    
    I1$_{but-CCH}$ & -79.13 & -76.41 & -78.04 & -75.92 & \\
    I2$_{but-CCH}$ & -77.06 & -73.96 & -75.88 & -72.26 & \\
    TS1$_{but-CCH}$ & -29.92 & -29.70 & -31.02 & -29.10 & \\
    TS2$_{but-CCH}$ & -26.17 & -23.40 & -25.55 & -22.68 & \\
    TS3$_{but-CCH}$ & -46.74 & -41.19 & -43.23 & -40.39 & \\ 
    \hline
  \end{tabular*}
\end{table*}

\begin{table*}
\small
  \caption{Relative energies (in kcal mol$^{-1}$) for the stationary points located along the gas-phase reaction paths of vinylacethylene with hydroxyl, cyano, and ethynyl radicals computed at different levels. ZPV energies included}
  \label{tbl:table4}
  \begin{tabular*}{\textwidth}{@{\extracolsep{\fill}}lrrrrr}
    \hline
    Molecule & M08HX & CC-F12//M08HX & B2PLYPD3 & CC-F12//B2PLD & CCSD(T)-F12 \\
    \hline
    \textit{Reaction \ch{H2CCHCCH} + \ch{OH}} & & & & & \\
    \ch{H2CCHCCH} + \ch{OH} & 0.00 & 0.00 & 0.00 & 0.00 & 0.00 \\
    \textit{syn}-\ch{H2CCHCCOH} + \ch{H} & 8.72 & 11.44 & 11.47 & 11.40 & 11.47 \\
    \textit{anti}-\ch{H2CCHCCOH} + H & 8.73 & 11.45 & 11.48 & 11.41 & 11.50 \\
    \textit{syn}-\ch{H2CCOHCCH} + \ch{H} & 1.91 & 2.74 & 3.36 & 2.47 & 2.59 \\
    \textit{anti}-\ch{H2CCOHCCH} + H & 0.95 & 1.87 & 2.55 & 1.71 & 1.80 \\
    \textit{syn-trans}-\ch{HOHCCHCCH} + \ch{H} & -0.61 & 1.03 & 1.19 & 0.86 & 0.99 \\
    \textit{anti-trans}-\ch{HOHCCHCCH} + H & 0.03 & 1.68 & 1.88 & 1.51 & 1.60 \\
    \textit{syn-cis}-\ch{HOHCCHCCH} + \ch{H} & -3.94 & -2.07 & -1.90 & -2.26 & -2.12 \\
    \textit{anti-cis}-\ch{HOHCCHCCH} + H & -0.06 & 1.44 & 2.21 & 1.87 & 1.96 \\    
    \textit{syn}-I1$_{vin-OH}$ & -37.94 & -35.56 & -34.78 & -35.46 & \\
    \textit{syn}-I2$_{vin-OH}$ &  -37.35 & -35.61 & -34.86 & -35.37 & \\
    \textit{syn}-TS1$_{vin-OH}$ & 3.81 & 4.95 & 5.11 & 5.10 & \\
    \textit{syn}-TS2$_{vin-OH}$ & 0.39 & 1.79 & 1.93 & 1.90 & \\
    \textit{syn}-TS3$_{vin-OH}$ & -36.60 & -35.00 & -34.11 & -34.77 & \\
    \textit{Reaction \ch{H2CCHCCH} + \ch{CN}} & & & & & \\
    \ch{H2CCHCCH} + \ch{CN} & 0.00 & 0.00 & 0.00 & 0.00 & 0.00 \\
    \ch{H2CCHCCCN} + \ch{H} & -28.47 & -20.75 & -26.09 & -20.52 & -20.60 \\
    \ch{H2CCCNCCH} + \ch{H} & -24.43 & -17.80 & -21.02 & -17.75 & -17.75 \\
    \textit{trans}-\ch{HCNCCHCCH} + \ch{H} & -28.24 & -20.80 & -24.60 & -20.65 & -20.67 \\
    \textit{cis}-\ch{HCNCCHCCH} + \ch{H} & -27.79 & -20.45 & -24.33 & -20.31 & -20.35 \\    
    I1$_{vin-CN}$ & -72.78 & -63.54 & -66.98 & -62.89 & \\
    I2$_{vin-CN}$ & -52.53 & -44.53 & -46.87 & -44.37 & \\
    I3$_{vin-CN}$ & -69.92 & -61.09 & -63.93 & -60.54 & \\
    I4$_{vin-CN}$ & -69.45 & -60.67 & -63.60 & -60.18 & \\
    TS1$_{vin-CN}$ & -23.14 & -15.88 & -19.80 & -15.00 & \\
    TS2$_{vin-CN}$ & -16.20 & -9.99 & -12.49 & -9.66 & \\
    TS3$_{vin-CN}$ & -24.18 & -17.03 & -20.09 & -16.43 & \\
    TS4$_{vin-CN}$ & -23.72 & -16.67 & -19.81 & -16.05 & \\
    TS5$_{vin-CN}$ & -9.76 & -4.51 & -9.10 & -4.34 & \\
    TS6$_{vin-CN}$ & -68.70 & -60.16 & -63.10 & -59.72 & \\
    \textit{Reaction \ch{H2CCHCCH} + \ch{CCH}} & & & & & \\ 
    \ch{H2CCHCCH} + \ch{CCH} & 0.00 & 0.00 & 0.00 & 0.00 & 0.00 \\
    \ch{H2CCHCCCCH} + \ch{H} & -34.30 & -30.05 & -35.59 & -29.92 & -29.71 \\
    \ch{H2CC(CCH)2} + \ch{H} & -27.64 & -24.89 & -28.12 & -24.85 & -24.52 \\
    \textit{trans}-\ch{HCCHCCHCCH} + \ch{H} & -31.42 & -27.73 & -31.65 & -27.64 & -27.37 \\
    \textit{cis}-\ch{HCCHCCHCCH} + \ch{H} & -31.09 & -27.48 & -31.42 & -27.31 & -27.07 \\    
    I1$_{vin-CCH}$ & -74.76 & -69.47 & -72.88 & -68.83 & \\
    I2$_{vin-CCH}$ & -54.23 & -50.25 & -52.54 & -50.13 & \\
    I3$_{vin-CCH}$ & -70.88 & -65.97 & -68.82 & -65.51 & \\
    I4$_{vin-CCH}$ & -70.23 & -65.39 & -68.33 & -64.97 & \\
    TS1$_{vin-CCH}$ & -28.73 & -25.03 & -29.03 & -24.17 & \\
    TS2$_{vin-CCH}$ & -19.30 & -17.09 & -19.53 & -16.79 & \\
    TS3$_{vin-CCH}$ & -27.35 & -24.06 & -27.09 & -23.46 & \\
    TS4$_{vin-CCH}$ & -26.96 & -23.74 & -26.87 & -23.10 & \\
    TS5$_{vin-CCH}$ & -4.56 & -3.38 & -7.49 & -2.98 & \\
    TS6$_{vin-CCH}$ & -69.46 & -64.86 & -67.75 & -64.45 & \\
    \hline
  \end{tabular*}
\end{table*}

\begin{figure*}
 \centering
 \includegraphics[height=8cm]{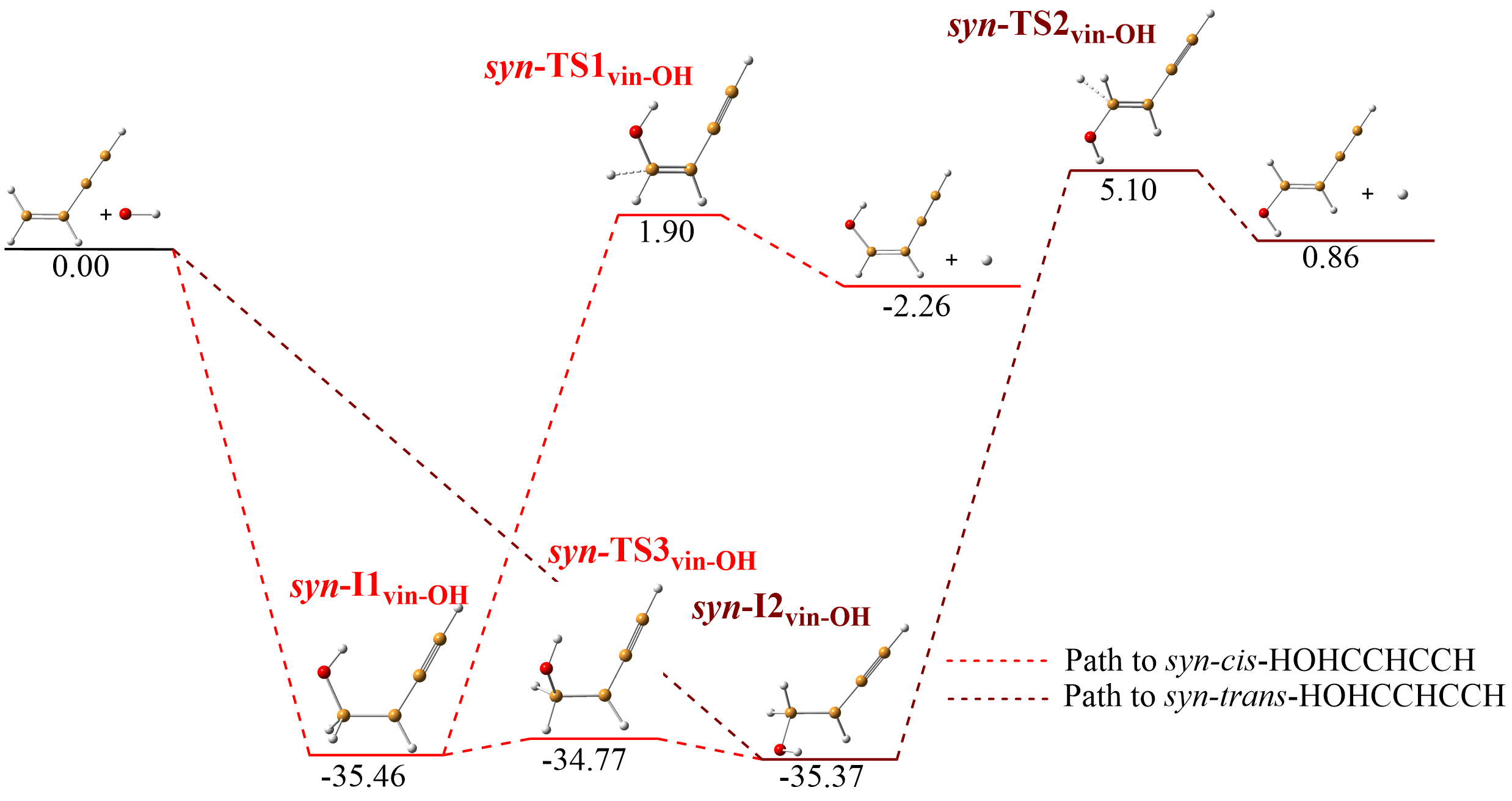}
 \caption{Relative energies (in kcal mol$^{-1}$) for the stationary points located along the gas phase reaction of vinylacetylene with hydroxyl radical computed at the CCSD(T)-F12/cc-pVTZ-F12//B2PLYPD3/aug-cc-pVTZ level. ZPV energies included.}
 \label{fgr:hccchch2+oh}
\end{figure*}

\subsubsection{Reaction of vinylacetylene with OH, CN, and CCH}

\begin{figure*}
 \centering
 \includegraphics[height=9.3cm]{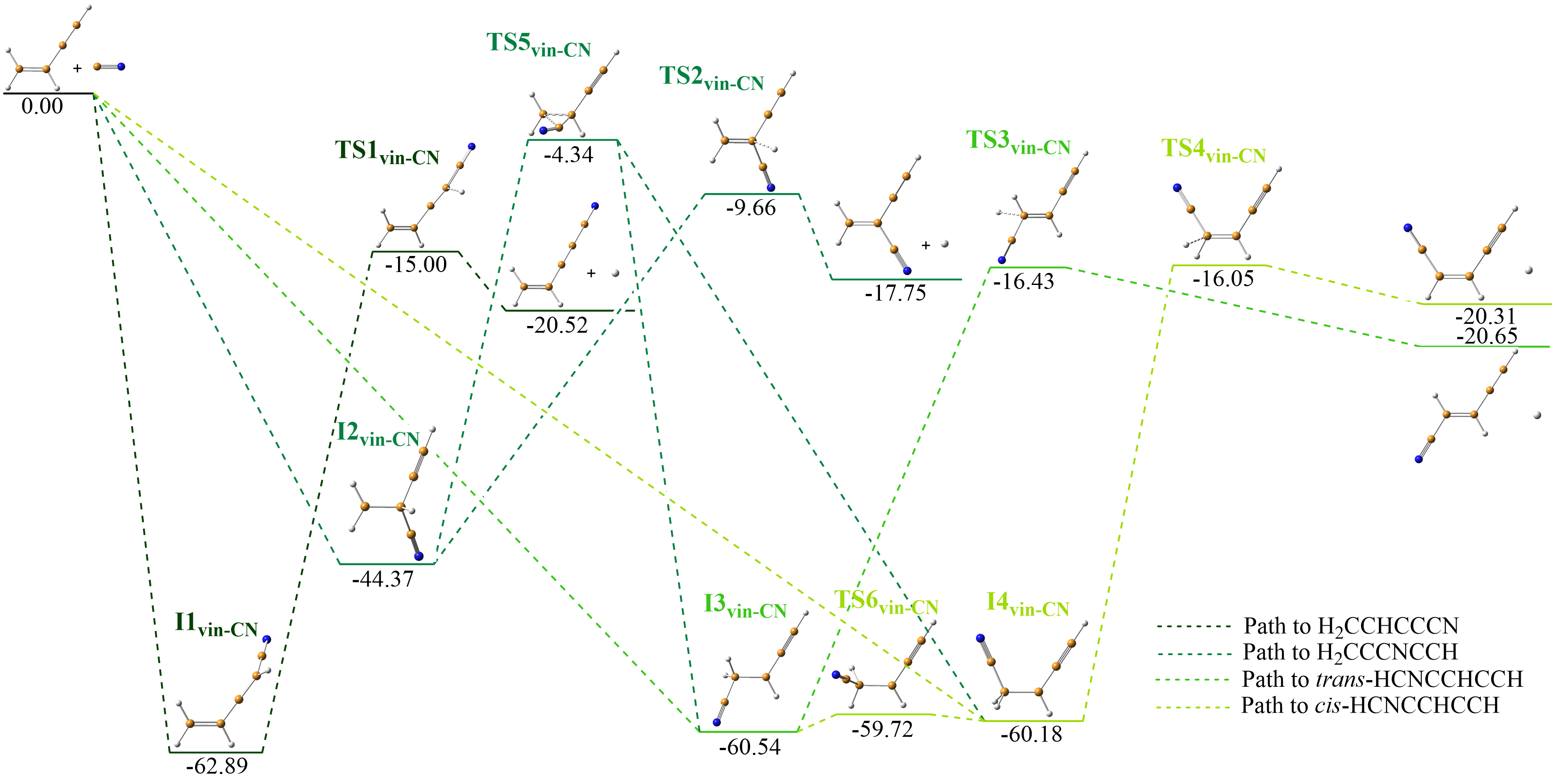}
 \caption{Relative energies (in kcal mol$^{-1}$) for the stationary points located along the gas phase reaction of vinylacetylene with cyano radical computed at the CCSD(T)-F12/cc-pVTZ-F12//B2PLYPD3/aug-cc-pVTZ level
 . ZPV energies included.}
 \label{fgr:hccchch2+cn}
\end{figure*}

\begin{figure*}
 \centering
 \includegraphics[height=9.5cm]{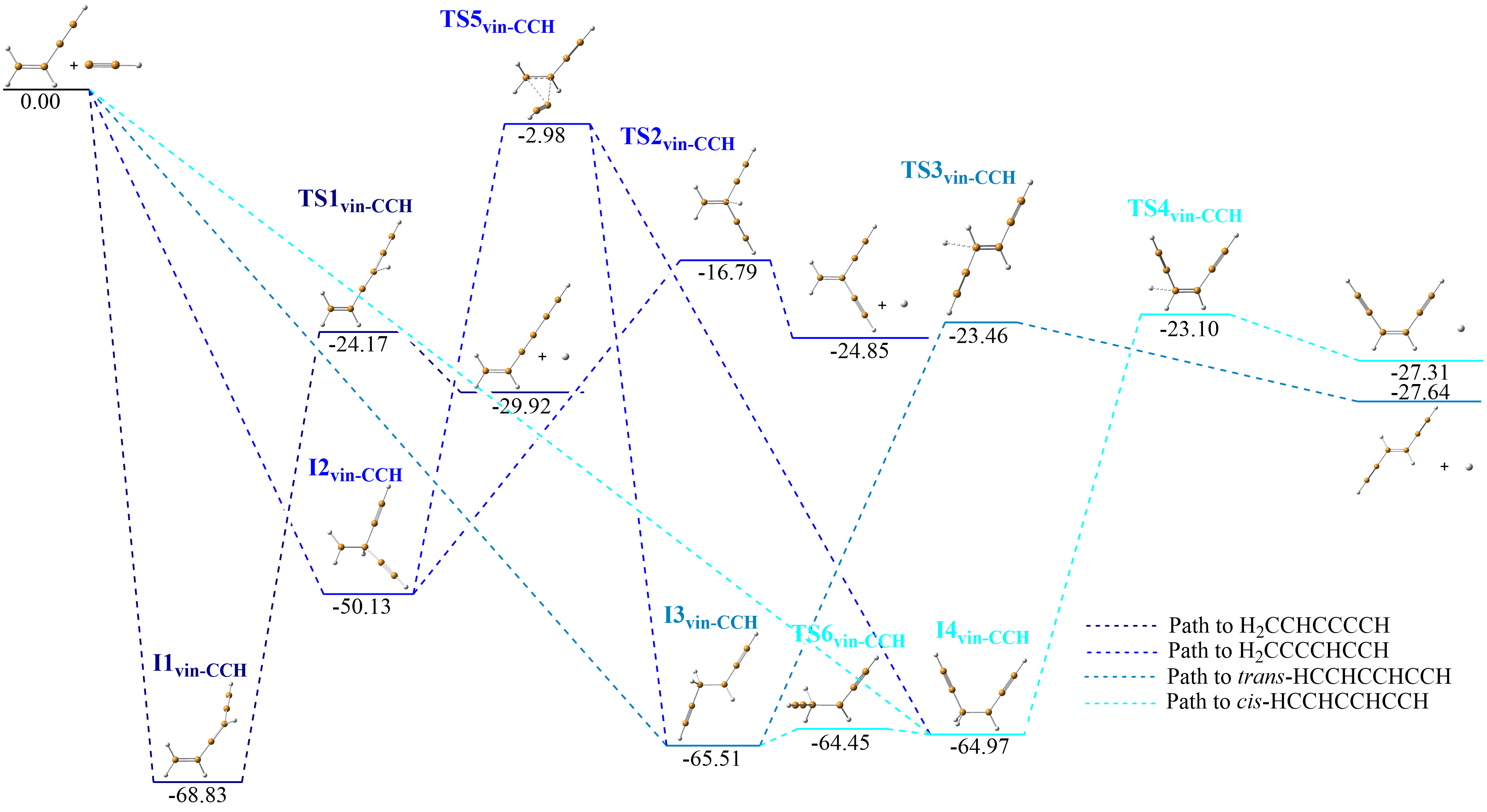}
 \caption{Relative energies (in kcal mol$^{-1}$) for the stationary points located along the gas phase reaction of vinylacetylene with ethynyl radical computed at the CCSD(T)-F12/cc-pVTZ-F12//B2PLYPD3/aug-cc-pVTZ level. ZPV energies included.}
 \label{fgr:hccchch2+cch}
\end{figure*}

Finally, we analyze the reactions of OH, CN, and CCH radicals with vinylacetylene. The relative energies of reactants, products, intermediates, and transition states characterized on their respective PESs -at the different calculation levels- are summarized in Table~\ref{tbl:table4}. The corresponding reaction profiles for OH, CN, and CCH are depicted in Fig.~\ref{fgr:hccchch2+oh}, ~\ref{fgr:hccchch2+cn}, and ~\ref{fgr:hccchch2+cch}, respectively.

We first consider the reaction of \ch{H2CCHCCH} with the OH radical. As can be seen in Table~\ref{tbl:table4}, the reaction is only exothermic for the formation of the vinylacetylene derivative, \textit{syn-cis}-\ch{HOHCCHCCH} with a reaction energy of -2.12 kcal mol$^{-1}$ at the CCSD(T)-F12 level. The reaction energy for obtaining the \textit{syn-trans}-\ch{HOHCCHCCH} isomer is close to zero at all the different levels employed. Therefore, we will consider the formation of these two products whose reaction profiles are shown in Fig.~\ref{fgr:hccchch2+oh}. When the OH radical interacts with the terminal carbon of the acetylene group, two intermediates, \textit{syn}-I1$_{vin-OH}$ and \textit{syn}-I2$_{vin-OH}$, are obtained, depending on the dihedral angle formed between the OH radical and the CCH group of vinylacetylene. The formation of these intermediates occurs directly, without the involvement of transition states and they are located below the reactants (-35.46, and -35.37 kcal mol$^{-1}$ at the CC-F12//B2PLD level, respectively). Subsequent hydrogen elimination from the OH-bonded carbon in \textit{syn}-I1$_{vin-OH}$ and \textit{syn}-I2$_{vin-OH}$ leads to the formation of the products \textit{syn-cis}-\ch{HOHCCHCCH} and \textit{syn-trans}-\ch{HOHCCHCCH}. These processes proceed through the transition states \textit{syn}-TS1$_{vin-OH}$ and \textit{syn}-TS2$_{vin-OH}$, which lie 1.90 and 5.10 kcal mol$^{-1}$ above the reactants at the CC-F12//B2PLD level, respectively.  Given these energy barriers, the formation of hydroxy derivatives of vinylacetylene through these reaction pathways are unlikely under the typical temperature and density conditions of the ISM. 

On the other hand, the formation pathways of the four cyano-substituted derivatives of vinylacetylene, \ch{H2CCHCCCN}, \ch{H2CCCNCCH}, \textit{trans}-\ch{HCNCCHCCH}, and \textit{cis}-\ch{HCNCCHCCH}, through the substitution of a hydrogen atom in vinylacetylene for the cyano (CN) radical, are exothermic, with reaction energies ranging from -17.75 to -20.67 kcal mol$^{-1}$, as summarized in Table~\ref{tbl:table4}. The corresponding reaction profiles for the formation of these cyano derivatives are depicted in Fig.~\ref{fgr:hccchch2+cn}.

The reaction initiates with the addition of the CN radical to one of the carbon atoms in vinylacetylene. When CN attaches to the carbon of the ethynyl moiety, intermediate I1$_{vin-CN}$ is formed. Alternatively, if the CN radical binds to the carbon of the vinyl group that is bonded to just one hydrogen, intermediate I2$_{vin-CN}$ is obtained. Furthermore, intermediates I3$_{vin-CN}$ and I4$_{vin-CN}$ arise from the attachment of CN to the terminal carbon of the vinyl group in different orientations. All intermediates lie significantly below the reactants, with relative energies of -62.89, -44.37, -60.54, and -60.18 kcal mol$^{-1}$ at the CC-F12//B2PLD level, respectively. From these intermediates, the elimination of a hydrogen atom bonded to the carbon where the CN radical was initially added results in the formation of the final cyano-substituted vinylacetylene products \ch{H2CCHCCCN}, \ch{H2CCCNCCH}, \textit{trans}-\ch{HCNCCHCCH}, and \textit{cis}-\ch{HCNCCHCCH}. These steps proceed through transition states TS1$_{vin-CN}$, TS2$_{vin-CN}$, TS3$_{vin-CN}$, and TS4$_{vin-CN}$, which are located -15.00, -9.66, and -16.43 and -16.05 kcal mol$^{-1}$ below the reactants at the CC-F12//B2PLD level, respectively. 
Additionally, we include in the reaction profile transition state 
TS5$_{vin-CN}$, located at -4.34 kcal mol$^{-1}$ relative to the reactants at the CC-F12//B2PLD level, which corresponds to the isomerization of intermediate I2$_{vin-CN}$ into the more stable ones I3$_{vin-CN}$ and I4$_{vin-CN}$. We also characterized the transition state TS6$_{vin-CN}$, which is associated with the rotation of the terminal vinyl carbon, facilitating the interconversion of I3$_{vin-CN}$ to I4$_{vin-CN}$. This transition state is positioned slightly above the intermediates, with a relative energy of -59.72 kcal mol$^{-1}$ at the CC-F12//B2PLD level.
As shown in the Fig.~\ref{fgr:hccchch2+cn}, the formation pathways of the four cyano-substituted vinylacetylene isomers follow analogous mechanisms, which can be exemplified by the formation of \ch{H2CCHCCCN}:

\small
\begin{equation}
\ch{H2CCHCCH} + \ch{CN} \rightarrow \mathrm{I1_{vin-CN}} \rightarrow \mathrm{TS1_{vin-CN}} \rightarrow \ch{H2CCHCCCN} + \ch{H}
\end{equation}
\normalsize

These results indicate that the formation of the four cyano-substituted derivatives of vinylacetylene is thermodynamically feasible in the ISM from the proposed CN addition. However, determining the relative abundances of each isomer requires a detailed kinetic study of the processes, which falls beyond the scope of this work and will be undertaken in the near future. As noted in the introduction, two cyanovinylacetylene isomers, \ch{H2CCHCCCN} and \textit{trans}-\ch{HCNCCHCCH}, have been detected in the TMC-1 molecular cloud.\citep{Cernicharo2021c,lee2021b} The reaction described here represent a potential synthetic pathway contributing to their formation.

As shown in Table~\ref{tbl:table4}, the formation of the four ethynyl-substituted derivatives of vinylacetylene, \ch{H2CCHCCCCH}, \ch{H2CC(CCH)2}, \textit{trans}-\ch{HCCHCCHCCH}, and \textit{cis}-\ch{HCCHCCHCCH}, from the reaction between vinylacetylene and the CCH radical is exothermic, with reaction energies of -29.71, -24.52, -27.37, and -27.07 kcal mol$^{-1}$ at the CCSD(T)-F12 level, respectively. The corresponding reaction pathways leading to these products are depicted in Fig.~\ref{fgr:hccchch2+cch}. It is worth noting that these pathways closely resemble those described in Fig.~\ref{fgr:hccchch2+cn} for the analogous reaction with the CN radical. The reaction between the CCH radical and vinylacetylene leads to the formation of four intermediates, I1$_{vin-CCH}$, I2$_{vin-CCH}$, I3$_{vin-CCH}$, and I4$_{vin-CCH}$. These intermediates undergo hydrogen atom elimination from the carbon where the CCH radical has bonded, resulting in the final products. The transition states involved in these processes, TS1$_{vin-CCH}$, 
TS2$_{vin-CCH}$, TS3$_{vin-CCH}$, and TS4$_{vin-CCH}$, are located well below the reactants (-24.17, -16.79, -23.46, and -23.10 kcal mol$^{-1}$ at the CC-F12//B2PLD level). Thus, the formation of isomers \ch{H2CCHCCCCH}, \ch{H2CC(CCH)2}, \textit{trans}-\ch{HCCHCCHCCH}, and \textit{cis}-\ch{HCCHCCHCCH}, from the CCH addition to vinylacethylene, is feasible under ISM conditions, as all processes proceed without a net activation barriers. In contrast, the higher TSs reported above for the reactions with the OH radicals can be rationalized in terms of a reduced delocalization of $\pi$-electrons in the double bonds, compared to CN- and CCH- bearing species, which benefit from greater stabilization due to a more extended conjugation.

\subsection{Astrophysical Implications}

Our theoretical findings demonstrate that the cyano, ethynyl, and hydroxy derivatives of the highly strained antiaromatic cycle, \ch{c-C4H4}, can readily form under ISM conditions, provided that the parental species \ch{c-C4H4} is available in the gas phase, as supported by its recent laboratory detection in low-temperature ice analogs.\cite{Wang2024} Subsequent desorption could occur through different thermal and non-thermal processes (e.g., in the hot-core stage or due to the action of large-scale shocks), depending on the physical properties of the astronomical environment. Furthermore, we emphasize that while \ch{c-C4H4} is “invisible” to radioastronomy due to its lack of a permanent dipole moment, the detection of its OH, CN and CCH derivatives could serve as indirect evidence of the presence of \ch{c-C4H4} in the ISM. Therefore, the present results highlight the need for future theoretical and laboratory studies on the spectroscopic characterization of these polar derivatives that facilitate their astronomical identification, which will help shed light on the role of antiaromatic systems in astrochemistry.

The results obtained for the reactions of butatriene with CN and CCH radicals suggest that the formation of butatriene derivatives, specifically \ch{H2C4HCN} and \ch{H2C4HCCH}, is energetically feasible. However, the formation of vinylacetylene derivatives, such as \ch{H2CCCNCCH} and \ch{H2CC(CCH)2}, is even more favorable, exhibiting higher exothermicity and lower activation barriers. In the case of the OH radical, only the formation of vinylacetylene derivatives, \textit{syn}- and \textit{anti}-\ch{H2CCOHCCH}, is found to be viable. On the basis of these results, a kinetic analysis is necessary to fully assess the viability of butatriene derivative formation under interstellar conditions.

Four distinct isomers can be obtained via the CN and CCH addition reaction to vinylacetylene: \ch{H2CCHCCCN}, \ch{H2CCCNCCH}, \textit{trans}-\ch{CNHCCHCCH}, and \textit{cis}-\ch{CNHCCHCCH}, \ch{H2CCHCCCCH}, \ch{H2CC(CCH)2}, \textit{trans}-\ch{HCCHCCHCCH}, and \textit{cis}-\ch{HCCHCCHCCH}. Bearing in mind that \ch{H2CCHCCCN} and \textit{trans}-\ch{HCNCCHCCH}, have been already detected in the cold dark cloud TMC-1,\citep{Cernicharo2021c,lee2021b} our results suggest that the CN and CCH addition reactions to vinylacetylene represent a plausible formation of pathway leading to these species. In this context, future theoretical efforts will focus on analyzing the kinetics behind the proposed routes, which will be essential for deriving and subsequently rationalizing the observed isomeric ratio.

Overall, our results indicate that cyano and ethynyl derivatives can be readily formed from \ch{C4H4} hydrocarbons. In contrast, the formation of hydroxy derivatives appears significantly less favorable, despite the high abundance of OH radicals in the interstellar medium (ISM). These results are consistent with the fact that, to date, only CN and CCH derivatives of pure cyclic hydrocarbons have been definitely identified.

\section{Conclusions}

In this study, we have performed a theoretical investigation on the stability and formation processes for the cyano (CN), ethynyl (CCH) and hydroxy (OH) derivatives of the \ch{C4H4} structural isomers: cyclobutadiene (\ch{c-C4H4}), butatriene (\ch{H2CCCCH2}), and vinylacetylene (\ch{H2CCHCCH}). Particularly, we proposed the gas-phase reaction: \ch{C4H4 + X → C4H3X + H}, where X = CN, CCH, and OH. For each system, we have analyzed in detail the thermochemistry and identified the critical points involved in each PES to determine possible activation barriers. The main conclusions of this work are the following:

\begin{itemize}

\item The results for the reaction starting with the cyclic system, cyclobutadiene (\ch{c-C4H4}), have shown that the formation of cyano, ethynyl, and hydroxy derivatives are thermodynamically feasible and involve no net activation barriers. Among these, the most favorable pathways involve the formation of \ch{c-C4H3CN} and \ch{c-C4H3CCH} derivatives, which emerge as promising candidates for interstellar detection via JWST (IR) observations or through radioastronomical observations, once rotational laboratory spectroscopic data become accessible. 

\item Regarding the formation of the butatriene  (\ch{H2CCCCH2}) derivatives, we found endothermic reaction pathways to the formation of the hydroxy derivatives, being therefore unfeasible under ISM conditions. However, the suggested PESs also allows the formation of various vinylacetylene derivatives \textit{syn}- and \textit{anti}-\ch{H2CCOHCCH}, which proceeds with no net activation barrier. On the contrary, the CN and CCH addition reactions are exothermic and lead to \ch{H2C4HCN} and \ch{H2C4HCCH} without net activation barriers. Interestingly, we also found that the formation of the vinylacetylene derivatives \ch{H2CCCNCCH} and \ch{H2CC(CCH)2} is even more favorable from the energetic (more exothermic) point of view and shows lower barriers. 

\item As for vinylacetylene (\ch{H2CCHCCH}), while the formation of the OH derivatives are again unlikely under the typical conditions of the ISM due to net activation barrier, we have identified feasible routes for the formation of all CN and CCH substituted derivatives. In this context, future theoretical efforts will focus on analyzing the kinetics behind the proposed routes. 

\item Finally, concerning the different computational levels employed in this study, the reaction energies calculated at the CCSD(T)-F12 level deviate by less than 1.0 kcal mol$^{-1}$ from those obtained at the CC-F12//B2PLD level, where single-point energy calculations were performed on geometries optimized at the B2PLYPD3 level. The results obtained with this approaches is highly consistent. Therefore, the influence of molecular geometry is comparatively less significant than the incorporation of electron correlation energy.

\end{itemize}

\section*{Author contributions}
PR and MSN conceived the project. PR and ML carried out the quantum-chemical computations. PR, and MSN drafted the manuscript. All authors discussed the results and reviewed the manuscript.
%We strongly encourage authors to include author contributions and recommend using \href{https://casrai.org/credit/}{CRediT} for standardised contribution descriptions. Please refer to our general \href{https://www.rsc.org/journals-books-databases/journal-authors-reviewers/author-responsibilities/}{author guidelines} for more information about authorship.

\section*{Conflicts of interest}
There are no conflicts to declare.

\section*{Data availability}

The data supporting this article have been included as part of the Supplementary Information. Additional data are available from the corresponding author upon reasonable request.
%A data availability statement (DAS) is required to be submitted alongside all articles. Please read our \href{https://www.rsc.org/journals-books-databases/author-and-reviewer-hub/authors-information/prepare-and-format/data-sharing/#dataavailabilitystatements}{full guidance on data availability statements} for more details and examples of suitable statements you can use.

\section*{Acknowledgments}

Financial support from the Spanish Ministerio de Ciencia e Innovación (PID2020-117742GB-I00/AEI/10.13039/501100011033) is gratefully acknowledged. M. S.-N. acknowledges a Juan de la Cierva Postdoctoral Fellow project JDC2022-048934-I, funded by the Spanish Ministry of Science, Innovation and Universities/State Agency of Research MICIU/AEI/10.13039/501100011033 and by the European Union “NextGenerationEU/PRTR”.

%%%END OF MAIN TEXT%%%


%\balance

%%%REFERENCES%%%
\bibliography{biblio,biblio2} 
\bibliographystyle{rsc} %the RSC's .bst file

\end{document}